\documentclass[a4paper,11pt]{article}
\usepackage{jheppub}
\usepackage{graphicx} %
\usepackage{color}
\usepackage[dvipsnames]{xcolor}
\usepackage[T1]{fontenc}
\usepackage{textcomp}
\usepackage{empheq}
\usepackage{hyperref}
\usepackage{amsmath} 
\usepackage{amssymb}
\usepackage[textsize=tiny]{todonotes}
\usepackage{comment}
\usepackage{multirow}
\usepackage{makecell}
\usepackage{booktabs}
\definecolor{cerulean}{rgb}{0., 0.52,0.65}

\newcommand{\bra}[1]{\left\langle #1 \right|}
\newcommand{\ket}[1]{\left|#1\right\rangle}

\newcommand{\bfr}{ {\bf r} }

\newcommand{\bfq}{ {\bf q} }
\newcommand{\bfv}{ {\bf v} }

\newcommand{\fancyO}{\mathcal{O}}
\newcommand{\fancyV}{\mathcal{V}}

\title{
Coherence from interference: a solvable model of sub-GeV dark matter-nucleus scattering
}
\author[a]{Lynn Lin}
\author[a]{Tongyan Lin}
\author[a]{Momei Fang}
\affiliation[a]{Department of Physics, University of California San Diego, La Jolla, CA 92093, USA}

\abstract{
    How do dark matter-nucleus interactions transition from the regimes of coherent scattering, where single phonons are produced, to that of individual nuclear recoils?
    Answering this question relies on understanding multiphonon excitations. Multiphonons are important for interpreting low-threshold direct detection experiments, yet are computationally prohibitive to compute. 
    In this paper, we employ a 1D $N$-site crystal lattice model where dark matter scattering can be computed exactly. We show that the only difference between coherent and incoherent scattering is that conservation of crystal momentum is enforced in coherent scattering. The momentum conservation constraint becomes less important as more phonons are produced, yielding the transition to incoherent scattering. 
    Using numerical calculations of the 1D structure factor, we also obtain quantitative validation of using an incoherent approximation to compute sub-GeV dark matter scattering in realistic 3D crystals. 
}

\begin{document}

\maketitle

\section{Introduction}

Over the past decades, dark matter (DM) direct detection has made significant progress in constraining particle masses $m_\chi>$ few GeV, where DM scattering produces nuclear recoils with $\gtrsim $ keV energies~\cite{LZ:2024zvo,XENON:2025vwd}. 
For DM masses below GeV (light DM), the recoil energies lie below such thresholds, and the parameter space is much less constrained. Theoretical and experimental progress has led to promising paths toward probing light DM~\cite{Kahn:2021ttr,Zurek:2024qfm}. For DM coupled to nucleons, recent experiments demonstrate eV-scale energy thresholds~\cite{TESSERACT:2025tfw, CRESST:2019jnq}, with the possibility of reaching DM down to keV masses via single-phonon detection~\cite{Knapen:2017ekk}.

Exploring DM direct detection in the keV-GeV mass range requires a fundamental shift in our theoretical modeling of scattering. For light DM, the momentum transfer $q$ and energy deposited $\omega$ become sufficiently small such that the approximation of elastic free nuclear recoils breaks down. In crystal targets, DM can instead scatter coherently from the lattice, producing collective excitations such as single phonons~\cite{Knapen:2017ekk,Griffin:2018bjn,Trickle:2019nya} or multiphonons~\cite{Knapen:2016cue,Campbell-Deem:2019hdx,Kahn:2020fef,Campbell-Deem:2022fqm,Gori:2025jzu}.  DM scattering into phonons in liquid helium targets has also been studied~\cite{Schutz:2016tid,Knapen:2016cue,Caputo:2019ywq,Acanfora:2019con,Caputo:2020sys}. Understanding scattering into phonons is also important for inelastic processes such as the Migdal effect~\cite{Knapen:2020aky,Berghaus:2022pbu,Berghaus:2026kmj}, which rely on charge excitations to detect low-energy recoils~\cite{Ibe:2017yqa}, and in predicting backgrounds~\cite{Berghaus:2021wrp}.

In the scattering rate, the DM-target cross-section is typically factorized to separate the unknown physics of DM particles from the target-specific physics capturing collective effects. This factorization is expressed in the differential cross-section as~\cite{Kahn:2021ttr,Campbell-Deem:2022fqm}:
\begin{equation}
    \frac{d\sigma}{d^3\bfq \, d\omega}=\frac{b^2_p}{\mu_\chi^2}\frac{1}{v}\frac{\Omega_c}{2\pi}|\tilde{F}(q)|^2S(\bfq, \omega)\delta(\omega-\omega_\bfq)
\end{equation}
In the prefactors, $b_p$ is the DM-proton scattering length, $\mu_\chi$ is the DM-proton reduced mass, $v$ is the initial velocity of the dark matter, and $\Omega_c$ is the unit cell volume.
The delta function captures the scattering kinematics through $\omega_\bfq = \bfq\cdot\bfv-q^2/2m_\chi$, where $\bfq$ is the momentum transferred to the crystal and $\omega$ is the energy transfer. The particle physics of the DM model is contained in the form factor $|\tilde{F}(q)|^2$. The target-specific physics is encapsulated by the dynamic structure factor, $S(\bfq,\omega)$. 
For a lattice with atoms labeled by $d$ in unit cells labeled by $\ell$, the structure factor can be written as
\begin{equation}
    S(\bfq, \omega) = \frac{1}{V}\sum_{ \ell,\ell'}\sum_{d,d'} f_d f_{d'} \int^\infty_{-\infty} dt \langle e^{-i\bfq\cdot  \bfr_{\ell'd'}(0)} e^{i\bfq\cdot  \bfr_{\ell d}(t)}\rangle e^{-i\omega t}.
    \label{eq:structure_factor_intro}
\end{equation}
where $V$ is the crystal volume, $\bfr_{\ell d}$ is the position vector of atom $d$ in unit cell $\ell$, and $f_d$ is an interaction strength between DM and atom $d$. For sub-GeV DM scattering, we require knowledge of the structure factor both in the low-$q$ coherent scattering regime, as well as the high $q$ regime of incoherent interactions off individual nuclei.

\begin{figure}
    \centering
    \includegraphics[width=0.49\linewidth]{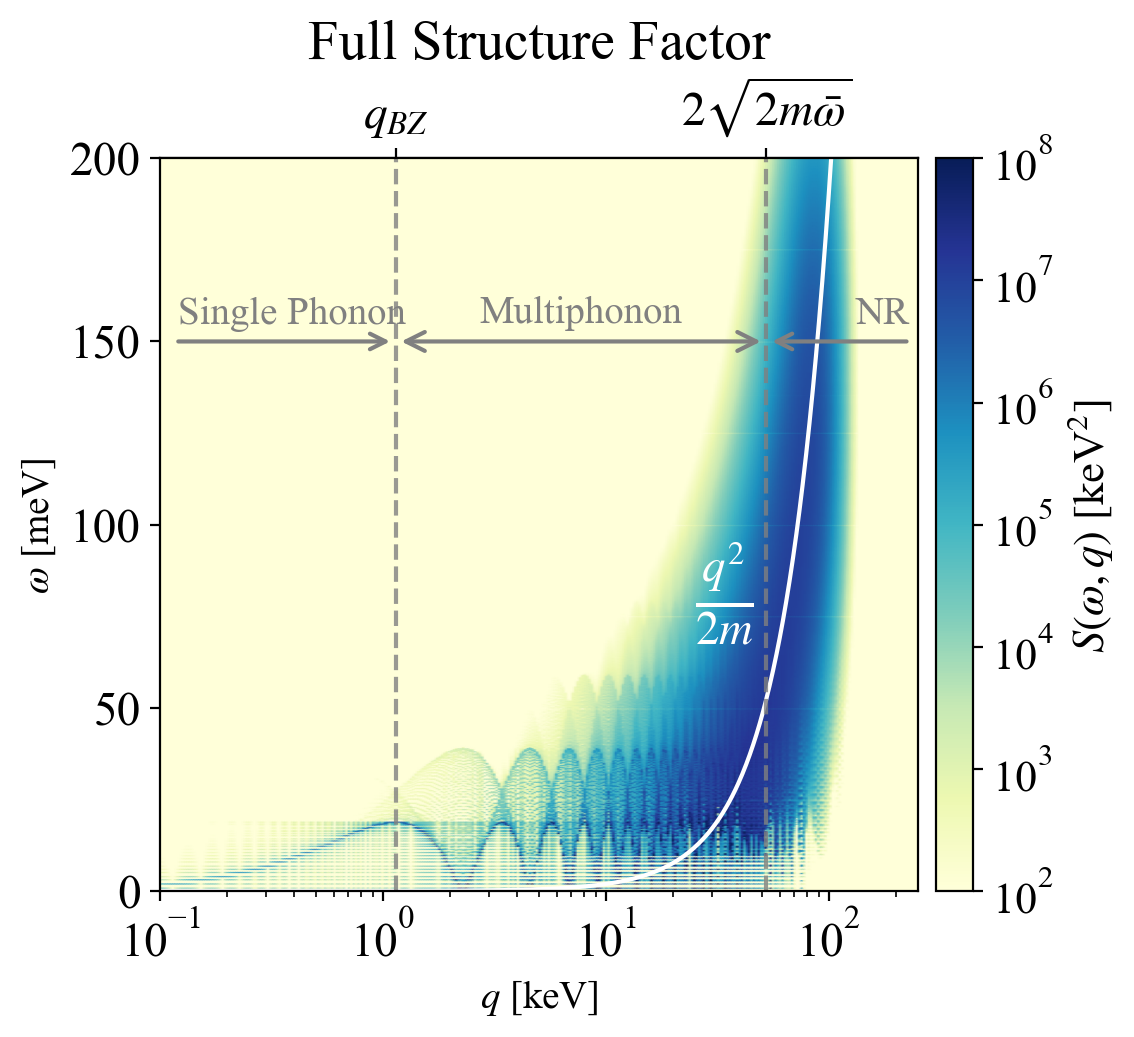}
    \includegraphics[width=0.49\linewidth]{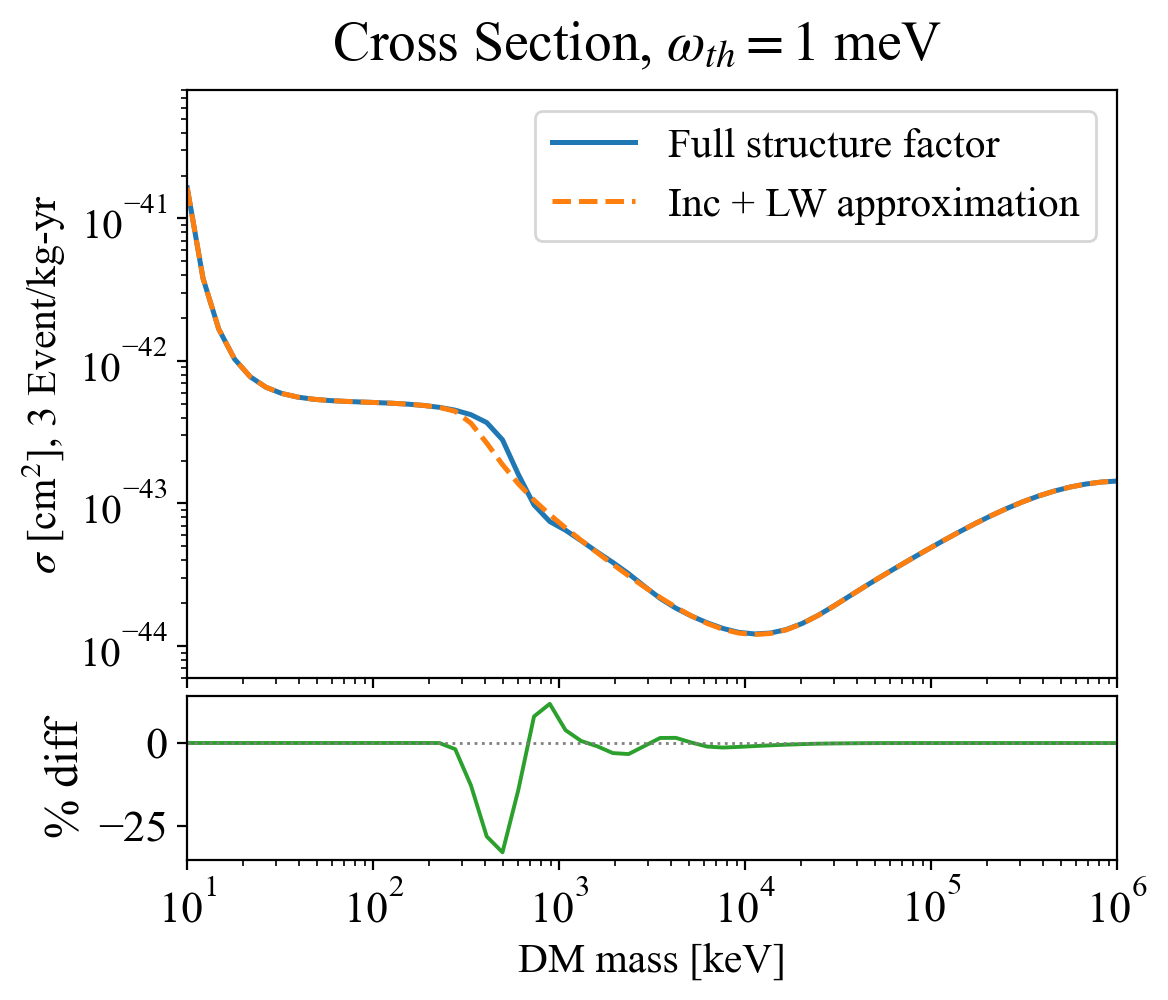}

    \caption{{\bf Left}: the full dynamic structure factor $S(q,\omega)$ in the 1D solvable model, evaluated across momentum transfer $q$ and energy transfer $\omega$. The plot illustrates the transition from the single-phonon regime ($q < q_{BZ} = \pi/a$) to the multiphonon and nuclear recoil (NR) regimes. The solid white curve indicates the standard nuclear recoil limit, $\omega = q^2/2m$. {\bf Right:} DM scattering cross-section as a function of DM mass for an energy threshold of 1 meV and 3 events/kg-year. The upper panel compares the exact full structure factor against the combined incoherent and long-wavelength (Inc+LW) approximations, and the lower panel shows the percentage difference between them. }
    \label{fig:sfactor-intro}
\end{figure}

For a crystal target, single phonon scattering is the dominant process for $q \lesssim \pi/a \sim O({\rm keV})$, where $a$ is the lattice spacing. The single phonon component of $S(\bfq,\omega)$ can be obtained exactly in terms of crystal properties such as phonon eigenvectors, and has been computed across a range of target materials~\cite{Knapen:2017ekk,Griffin:2018bjn,Cox:2019cod,Trickle:2019nya,Griffin:2019mvc,Caputo:2019xum,Griffin:2020lgd,Trickle:2020oki,Taufertshofer:2023rgq,Ashour:2024xfp,Li:2026xgj}. The potential for single phonon scattering to give daily modulation effects in anisotropic materials~\cite{Griffin:2018bjn,Coskuner:2021qxo,Griffin:2020lgd,Taufertshofer:2023rgq} has also been explored. Reproducing this single phonon behavior relies essentially on the interference terms in Eqn.~\eqref{eq:structure_factor_intro}.

For larger $q \sim \sqrt{2 m_N \bar \omega} \sim 50$ keV, with $m_N$ nucleus mass and $\bar \omega$ average phonon energy, multiphonon excitations become increasingly important in $S(\bfq,\omega)$, and for $q \gg \sqrt{2 m_N \bar \omega}$ the structure factor transitions to the expected behavior in the nuclear recoil regime, $S(\bfq,\omega) \propto \delta( \omega - q^2/(2m_N))$. However, generalizing single-phonon calculations to multi-phonon excitations is much more expensive computationally. Thus far, calculations of multi-phonon excitations and the nuclear recoil limit have relied on a variety of simplifying assumptions~\cite{Knapen:2016cue,Campbell-Deem:2019hdx,Kahn:2020fef,Knapen:2021bwg,Campbell-Deem:2022fqm,Lin:2023slv,Stratman:2024sng}. In particular, Ref.~\cite{Campbell-Deem:2022fqm} applied the {\emph{incoherent approximation}}, which drops all interference terms (cross terms) with $\ell\neq\ell'$ and $d\neq d'$ in Eqn.~\eqref{eq:structure_factor_intro}, for $q \gtrsim O({\rm keV})$. This effectively treats the crystal as a collection of independent oscillators rather than as a coupled system.  With this simplification, multiphonon excitations can be computed in terms of the phonon density of states and the nuclear recoil limit can be obtained.

While the incoherent approximation provides a computationally inexpensive way to model the structure factor, prior work did not address the accuracy of this approximation, particularly at low $q$.  In order to combine the incoherent approximation with the explicit calculations of single-phonon scattering, a more robust understanding of the transition from coherent to incoherent scattering is needed.

In this paper, we address this gap by employing a 1D $N$-site lattice model, which enables an exact analytic calculation of the full structure factor. With this model, we directly probe the importance of interference effects, relevant for coherent scattering, and identify the regimes where scattering is well-modeled by the incoherent approximation. We show explicitly that the only difference between incoherent and coherent scattering is the appearance of a delta-function that conserves crystal momentum. The conservation of crystal momentum reflects the discrete translation symmetry of the full crystal, while incoherent scattering treats atoms as individual scattering centers, breaking the symmetry. This intuition for the difference between incoherent and coherent scattering thus generalizes beyond 1D to 3D crystals as well.

Our numerical result for the full 1D structure factor is shown in the left panel of Fig. \ref{fig:sfactor-intro}. In the low $q$ regime, the structure factor of the 1D $N$-site lattice reduces to a Dirac delta function centered at the dispersion relation in the continuum limit. This reflects the constraints of energy and momentum conservation. At larger $q$ and $\omega$, the multiphonon structure factor still displays vestiges of the single-phonon dispersion and the kinematic constraints. However, in the scattering rate spectrum $dR/d\omega$ relevant for experiments, the incoherent approximation provides a reliable and valid approximation for $n \ge 2$ phonons. The right panel shows the DM cross section needed for 3 events/kg-year. A hybrid approximation scheme (Inc + LW) that integrates the incoherent approximation (applicable for $n\ge 2$ phonons or $q > $ keV) with coherent single-phonon scattering (relevant for $q < $ keV) yields  accurate results over most of the mass range. For $\omega_{th}=1$ meV, the maximum error does not exceed 30\% in the case of a massive mediator, and is percent-level or less outside of the 0.2$-$1 MeV mass window.

Sec.~\ref{sec:structure_factor} reviews the structure factor and defines the incoherent approximation. In Sec.~\ref{sec:2site}, we work through a two-site toy model where interference terms are visible in closed form. In Sec.~\ref{sec:Nsite}, we solve the $N$-site lattice exactly, show that the interference terms are precisely what enforce conservation of crystal momentum, and quantify where the incoherent approximation succeeds and fails. In Sec.~\ref{sec:scattering_rate} we compute scattering rates and validate the combined Inc + LW scheme, and in Sec.~\ref{sec:conclusions} we conclude. App.~\ref{app:recursive-relation} gives the recursion relations used for the numerics and App.~\ref{app:massless_mediator} collects additional massless mediator results.

\section{Structure Factor}
\label{sec:structure_factor}

Here we provide a brief review of some basic concepts of the structure factor~\cite{Squires1996,Schober2014}. In this paper, we focus on a simple 1D crystal model with one atom per unit cell and present a simplified expression for the structure factor. For a general 3D model, see~\cite{Campbell-Deem:2022fqm,Kahn:2021ttr}. 

To begin with, we consider the potential of an $N$-site lattice whose atom is labeled with $\ell$. For a DM particle with momentum transfer $q$, the Fourier space potential is
\begin{equation}
  \tilde{\fancyV}(q)=\frac{2\pi b_p}{\mu_\chi}\tilde{F}(q)\sum_{\ell}^N f_\ell e^{i q r_\ell}\equiv \frac{2\pi b_p}{\mu_\chi}\tilde{F}(q)\fancyO_T(q)
\end{equation}
where $r_\ell$ is the position of the $\ell$th atom, $b_p$ is the DM-proton scattering length defined by $\sigma_p\equiv 4\pi b_p^2$ at some reference momentum, and $\mu_\chi$ is the DM-proton reduced mass. For a heavy mediator, the form factor $\tilde{F}(q)=1$, and for a massless mediator, $\tilde{F}(q)=q_0^2/q^2$. The coupling constant $f_\ell$ describes the effective coupling to the atom $\ell$; in our simplified model with one atom per unit cell, $f_\ell = f$.

The operator $\fancyO_T(q)\equiv\sum_{l}^N f e^{i {q}{r}_\ell}$ acts only on the target lattice, and can therefore be factored out in our analysis. We define the dynamic structure factor as
\begin{equation}
    S(q, \omega)\equiv \frac{2\pi}{V}\sum_f\left|\langle \Phi_f|\fancyO_T(\textbf{q})|0\rangle \right|^2 \delta(E_f-\omega)= \frac{2\pi}{V}\sum_f\left|\sum^N_\ell \langle \Phi_f|f e^{i qr_\ell}|0\rangle \right|^2 \delta(E_f-\omega).
\end{equation}
We assume the system is excited from the ground state $\ket{0}$ to the final state $\ket{\Phi_f}$ by the DM collision, and all possible final states $f$ are then summed over to get the full structure factor. The delta function enforces the conservation of energy. The structure factor describes the target's response to the DM as a function of the energy deposit $\omega$ and momentum transfer $q$. 

We can rewrite the structure factor by expanding the square and Fourier transforming the $\delta$-function, yielding  
\begin{equation}
    S({q},\omega) = f^2 \sum_{\ell,\ell'}^N  C_{{\ell}{\ell'}},
\end{equation}
where the correlation function $C_{\ell\ell'}$ is defined as 
\begin{equation}
\begin{split}
    C_{\ell\ell'}&\equiv\frac{1}{V}\int^\infty_{-\infty}dt\sum_f\bra{0}e^{-iq r_{\ell'}(0)}\ket{\Phi_f}\times\bra{\Phi_f}e^{iq r_\ell(t)}\ket{0}e^{-i\omega t}\\
    &\equiv \frac{1}{V}\int^\infty_{-\infty} dt \langle e^{-iq r_{\ell'}(0)} e^{iq r_\ell(t)}\rangle e^{-i\omega t}.
\end{split}
\end{equation}
This  yields a structure factor with form equivalent to Eqn.~\eqref{eq:structure_factor_intro}.

To evaluate the correlation function, we express the atomic position $r_\ell$ as ${l}_\ell+u_\ell$, where the lattice vector $|{l}_\ell| = a\ell$ in 1D space, and $u_\ell$ is the displacement vector, whose expansion is
\begin{equation}
    u_\ell(t) = \sum_\nu \frac{1}{\sqrt{2Nm\omega_\nu}}\left(
        \textbf{e}_{\nu,\ell}\hat{a}_\nu e^{-i\omega_\nu t}  +\textbf{e}_{\nu,\ell}^*\hat{a}_\nu^\dagger e^{i\omega_\nu t} 
        \right)
\end{equation}
where $\nu$ labels the phonon branches, $\hat{a}_\nu^\dagger$ and $\hat{a}_\nu$ are creation and annihilation operators, $\omega_\nu$ is the phonon energy, and $\textbf{e}_{\nu,\ell}$ is the phonon eigenvector of branch $\nu$. We then apply the Baker-Campbell-Hausdorff formula, truncate all terms up to the first commutator, and then use Bloch's identity $\langle e^{\hat{A}} \rangle=e^{\frac{1}{2} \langle \hat{A}^2 \rangle}$ to get 
\begin{equation}
    C_{{\ell }{\ell'} } = \frac{1}{V}e^{iqa(\ell-\ell')} \int_{-\infty}^{\infty} dt \, e^{-\frac{1}{2}\langle ({q}  {u_{\ell '}}(0))^2\rangle} e^{-\frac{1}{2}\langle ({q} {u_{\ell}}(t))^2\rangle} e^{\langle {q}  {u_{\ell '}}(0)\, {q} {u_{\ell}}(t) \rangle}e^{-i\omega t}
\end{equation}
Since $\langle ({q} \, {u_\ell}(t))^2 \rangle$ is independent of time and lattice site, we define the Debye-Waller factor as $\exp{(-2W(q))}$ with 
\begin{equation}
    W({q})\equiv \frac{1}{2}\langle ({q} {u_\ell}(0))^2 \rangle. \label{eqn:DW-def}
\end{equation}
We use the previous equation to rewrite $C_{{\ell}{\ell'}}$ as
\begin{equation}
         C_{{\ell}{\ell'} } = \frac{1}{V}e^{i{q}a(\ell-{\ell'})} e^{-2W(q)} \int_{-\infty}^{\infty} dt \, e^{\langle {q}  {u_{\ell '}}(0)\, {q}  {u_{\ell}}(t) \rangle}e^{-i\omega t} \label{eqn:cll-def}
\end{equation}
In the following sections, we will explicitly evaluate Eqn. \eqref{eqn:cll-def} to calculate the structure factor of different models. Note that we work in the harmonic limit throughout; the effect of anharmonic interactions is estimated to be small in most parts of phase space~\cite{Lin:2023slv}.

In the incoherent approximation, the interparticle interference terms are dropped: %
\begin{equation}\label{eqn:incoh}
    S_\text{inc}(q,\omega) \approx f^2\sum_{\ell}^N C_{\ell\ell}.
\end{equation}
The intuition is that the cross terms carry phases $e^{i{q}a(\ell-{\ell'})}$ which are rapidly oscillating across different lattice separations and cancel in the sum.
The auto-correlation function $C_{\ell \ell}$ can be computed in terms of multi-phonon excitations~\cite{Campbell-Deem:2022fqm}:
\begin{equation}
    S_\text{inc}(q, \omega)\approx \frac{2\pi{f}^2}{\Omega_c} e^{-2W(q)}\sum_n \frac{1}{n!}\left(\frac{q^2}{2m} \right)^n \left(\prod_{i=1}^{n}\int d\omega_i\frac{D(\omega_i)}{\omega_i}\right)\delta\left(\sum_j\omega_j-\omega\right). \label{eqn:s-dos}
\end{equation}
where we used that $V= N \Omega_c$ for unit cell volume $\Omega_c$. This form can be used to estimate that a given $n$-phonon term is most relevant at momentum scales $q \sim \sqrt{n} \sqrt{2 m \bar \omega}$ with $\bar \omega = \int d\omega \, \omega D(\omega)$. 
When $q \gg \sqrt{2 m \bar \omega}$, the structure factor can be described by the impulse approximation. 
Mathematically, this profile can be derived from a steepest descent expansion about $t=0$ in the correlation function in Eqn. \eqref{eqn:cll-def}, yielding 
\begin{equation}
    S_{IA} (q, \omega) = \frac{f^2}{\Omega_c}\sqrt{\frac{2\pi}{\Delta^2}}\exp\left( -\frac{(\omega-\frac{q^2}{2m})^2}{2\Delta^2}\right)
    \label{eq:S_IA}
\end{equation}
where $\Delta^2 = \frac{q^2\bar\omega}{2m}$. Physically, the impulse approximation treats the target as an effectively free, independent ion because the collision occurs too rapidly for the nucleus to interact with its neighbors. It is a continuous Gaussian distribution that peaks at the free nuclear recoil energy, $\omega=q^2/2m$.

\section{Toy Model: 2 sites} \label{sec:2site}

\begin{figure}[h]
    \centering
    \begin{tabular}{cc}
        \includegraphics[width=0.4\linewidth]{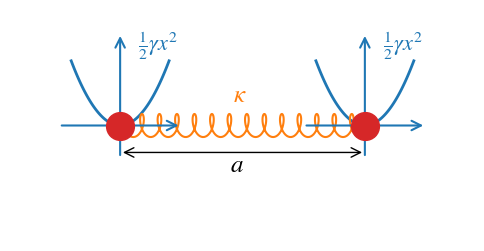} & 
        \includegraphics[width=0.4\linewidth]{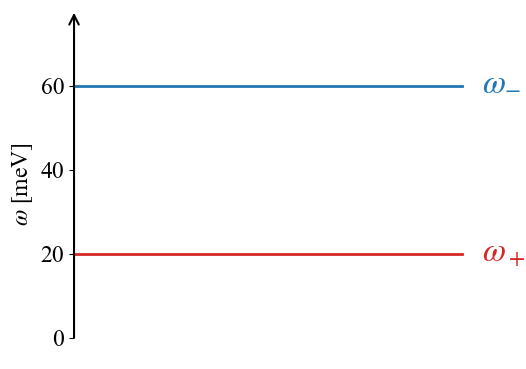} \\
        (a) & (b) %
    \end{tabular}
    \caption{Setup and energy spectrum for the two-site toy model. (a) Schematic representation of the system, where two adjacent atoms of mass $m$ separated by lattice spacing $a$ are confined by local harmonic potentials (stiffness $\gamma$) and coupled by a central spring (stiffness $\kappa$). (b) The corresponding normal mode energies, chosen as $\omega_+$ = 20 meV and $\omega_-$ = 60 meV, approximate the acoustic and optical phonon branches of silicon. }\label{fig:2site-schematic}
\end{figure}

Much of the essential physics can be clearly seen in the two-site model. A schematic diagram of the model is provided in Fig. \ref{fig:2site-schematic}(a). We assume a linear elastic response where the forces depend on both the individual particle displacements, $u_1$, $u_2$, and their relative displacement, $|u_1-u_2|$. We model this system using particles with mass $m$, momenta $p_1$, $p_2$, and lattice spacing $a$. Both atoms are confined by a harmonic potential and connected together by a spring of stiffness $\kappa$. 

The Hamiltonian of the system is
\begin{equation}
\begin{split}
        \hat{H} &= \frac{1}{2m}(p_1^2+p_2^2)+\frac{\gamma}{2}(u_1^2+u_2^2)+\frac{\kappa}{2}(u_1-u_2)^2\\  &\equiv \frac{1}{2m}(p_+^2+p_-^2)+\frac{\gamma}{2}(u_+^2+u_-^2)+\kappa u_-^2
\end{split}
\end{equation}
The normal modes of the system are $u_+=\frac{u_1+u_2}{\sqrt{2}}$, $u_-=\frac{u_1-u_2}{\sqrt{2}}$, with corresponding momenta $p_+ = \frac{p_1+p_2}{\sqrt{2}}$, $p_-=\frac{p_1-p_2}{\sqrt{2}}$, and energies $\omega_+ = \sqrt{\frac{\gamma}{m}}$ and $\omega_-=\sqrt{\frac{\gamma+2\kappa}{m}}$. As illustrated in Fig. \ref{fig:2site-schematic}(b), we choose $\omega_+=20$ meV and $\omega_-=60$ meV, corresponding to the acoustic and optical phonon energies of silicon, so the calculated structure factor reflects experimentally relevant energy scales.
The decoupled harmonic oscillators thus have displacements given by:
\begin{equation} 
    u_\pm = \sqrt{\frac{1}{2m\omega_\pm}}(\hat a_\pm^\dagger e^{i\omega_\pm t} + \hat a_\pm e^{-i\omega_\pm t})\label{eqn:2site-displacement}
\end{equation}

Using Eqn. \eqref{eqn:DW-def} and Eqn. \eqref{eqn:cll-def}, we can compute the Debye-Waller factor and the correlation functions: 
\begin{align}
    W(q)& = \frac{q^2}{8m}\left(\frac{1}{\omega_+}+\frac{1}{\omega_-}\right)\\
    C_{11}=C_{22} &= \frac{1}{V} e^{-2W(\boldsymbol{q})} \sum_{n=0} \left(\frac{q^2}{4m}\right)^n \sum_{j=0}^n \frac{2\pi}{j!(n-j)!\omega_+^j \omega_-^{n-j}} \delta \big ( j\omega_+   + (n-j) \omega_-  -\omega \big)\\
    C_{12}=C_{21} &= \frac{1}{V}e^{iqa} e^{-2W(\boldsymbol{q})} \sum_{n=0}\left(\frac{q^2}{4m}\right)^n \sum_{j=0}^n \frac{2\pi(-1)^{n-j}}{j!(n-j)!\omega_+^j \omega_-^{n-j}} \delta \big( j\omega_+  + (n-j) \omega_- - \omega \big)
\end{align}
where $n$ represents the total number of phonons, while $j$ and $n-j$ represent the number of excitations (phonons) in $\omega_+$ and $\omega_-$ states, respectively. 

The two correlation functions play distinct physical roles. $C_{11}$ can be understood as the energy-deposition spectrum when a momentum $q$ is imparted to a single atom. Summing $C_{11}$ and $C_{22}$ reproduces the incoherent approximation of Eq. \eqref{eqn:incoh}.
The cross term $C_{12}$ instead encodes coherence between the excitation amplitudes at the two sites.

Taking $n=1$ and summing $C_{{\ell}{\ell'}}^{(1)}$ gives the first-order structure factor and the incoherent approximation:
\begin{align}
    S^{(1)}_\text{full} &= \frac{4\pi f^2}{V}e^{-2W(q)} \frac{q^2}{4m}\left(\frac{1+\cos(qa)}{\omega_+}\delta(\omega-\omega_+)+\frac{1-\cos(qa)}{\omega_-}\delta(\omega-\omega_-)\right), \\
    S^{(1)}_{\text{inc}} &= \frac{4\pi f^2}{V}e^{-2W(q)} \frac{q^2}{4m}\left(\frac{1}{\omega_+}\delta(\omega-\omega_+)+\frac{1}{\omega_-}\delta(\omega-\omega_-)\right).
\end{align}
Relative to the incoherent approximation, the full structure factor contains additional $\cos(qa)$ terms, which qualitatively modify the structure factor at low $q$: 
\begin{equation}
    S^{(1)}_\text{full}\approx \frac{4\pi f^2}{V} \left(  \frac{q^2}{2m \omega_+}\delta(\omega-\omega_+) + \frac{q^4 a^2}{8 m \omega_-}\delta(\omega-\omega_-) \right).
\end{equation}
The interference between the two sites suppresses DM scattering into the optical mode $\omega_-$ by $q^2 a^2$~\cite{Griffin:2018bjn,Cox:2019cod}. Physically, at the $q\to 0$ limit, the DM can only drive their center-of-mass motion, not their relative displacement; scattering into the optical branch is thus forbidden by symmetry in this limit. This matches the standard long-wavelength approximation for single-phonon scattering. 

\begin{figure}[t]
    \centering
    \includegraphics[width=0.95\linewidth]{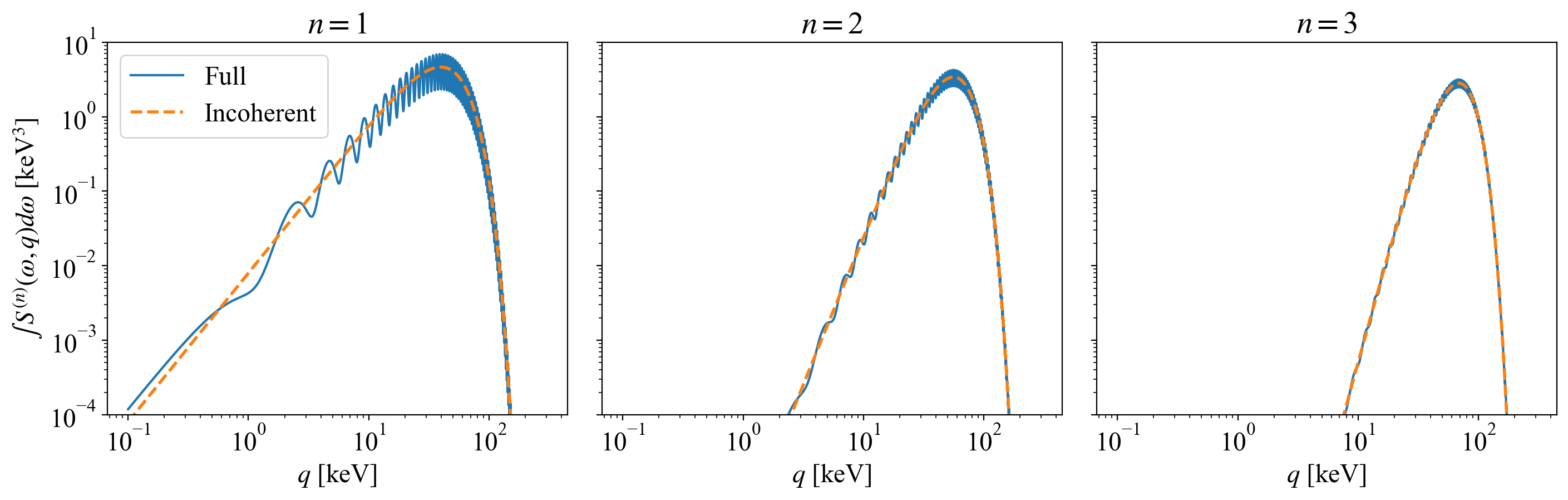}
        \caption{The energy-integrated structure factor, $\int S^{(n)}(q,\omega)d\omega$, plotted as a function of momentum transfer $q$ for the 1D two-site lattice model. The three panels display contributions from single-phonon ($n=1$), two-phonon ($n=2$), and three-phonon ($n=3$) processes. The full structure factor (solid line) exhibits prominent oscillations at $n=1$ due to inter-particle interference. However, these oscillations rapidly diminish at higher phonon orders ($n=2, 3$) due to phase cancellation across different terms in the structure factor, causing the full structure factor and the incoherent approximation to converge.}\label{fig:2state-int}
\end{figure}

Meanwhile, at high $q \gg \pi /a$ these interference terms cause rapid oscillations. When calculating a scattering rate and integrating over a wide range of $q$, these oscillations would average to zero, giving back the result from using the incoherent approximation. Fig.~\ref{fig:2state-int} compares the full structure factor and incoherent approximation as a function of $q$. Because the structure factor consists of delta functions, we integrate $S$ over $\omega$ and display $\int S^{(n)}(\omega,q) d\omega$ in  Fig.~\ref{fig:2state-int}.

The next few orders of the full structure factor are shown below:
\begin{align}
\begin{split}
    S^{(2)}_\text{full} &= \frac{4\pi f^2}{V}e^{-2W(q)} \left(\frac{q^2}{4m}\right)^2\left( \frac{1+\cos(qa)}{2\omega_+^2} \delta(\omega-2\omega_+)+\frac{1-\cos(qa)}{\omega_+\omega_-} \delta(\omega-\omega_+-\omega_-)\right. \\ 
    &\hspace{4.5cm}\left. +\frac{1+\cos(qa)}{2\omega_-^2} \delta(\omega-2\omega_-)\right)
\end{split}
\label{eq:2site_n2}\\
\begin{split}
    S^{(3)}_\text{full} &= \frac{4\pi f^2}{V}e^{-2W(q)} \left(\frac{q^2}{4m}\right)^3\left( \frac{1+\cos(qa)}{6\omega_+^3}\delta(\omega-3\omega_+)+\frac{1-\cos(qa)}{2\omega_+^2\omega_-}\delta(\omega-2\omega_+-\omega_-)\right.\\
    &\left.\hspace{2.5cm}+\frac{1+\cos(qa)}{2\omega_+\omega_-^2}\delta(\omega-\omega_+-2\omega_-)+\frac{1-\cos(qa)}{6\omega_-^3} \delta(\omega-3\omega_-)\right)
\end{split}
\label{eq:2site_n3}
\end{align}
These higher order structure factors become increasingly important at large $q$. Each individual delta-function still carries a $\cos(qa)$ interference term, which gives similar effects as in the $n=1$ case. However, in this case, these interference terms will be further suppressed when a range of energies is integrated over. This is shown in Fig. \ref{fig:2state-int}, when the amplitude of the oscillations in $q$ decreases with increasing $n$. This damping occurs because the interference terms include an alternating phase factor of $(-1)^{n-j}$, as can be seen in the coefficients of the $\cos(qa)$ terms in Eqns.~(\ref{eq:2site_n2}-\ref{eq:2site_n3}).  This leads to a cancellation at higher orders when integrating over a range of energies. Thus, the incoherent approximation becomes an increasingly good approximation with larger $n$.

\begin{figure*}[t]
    \centering
    \includegraphics[width=0.95\linewidth]{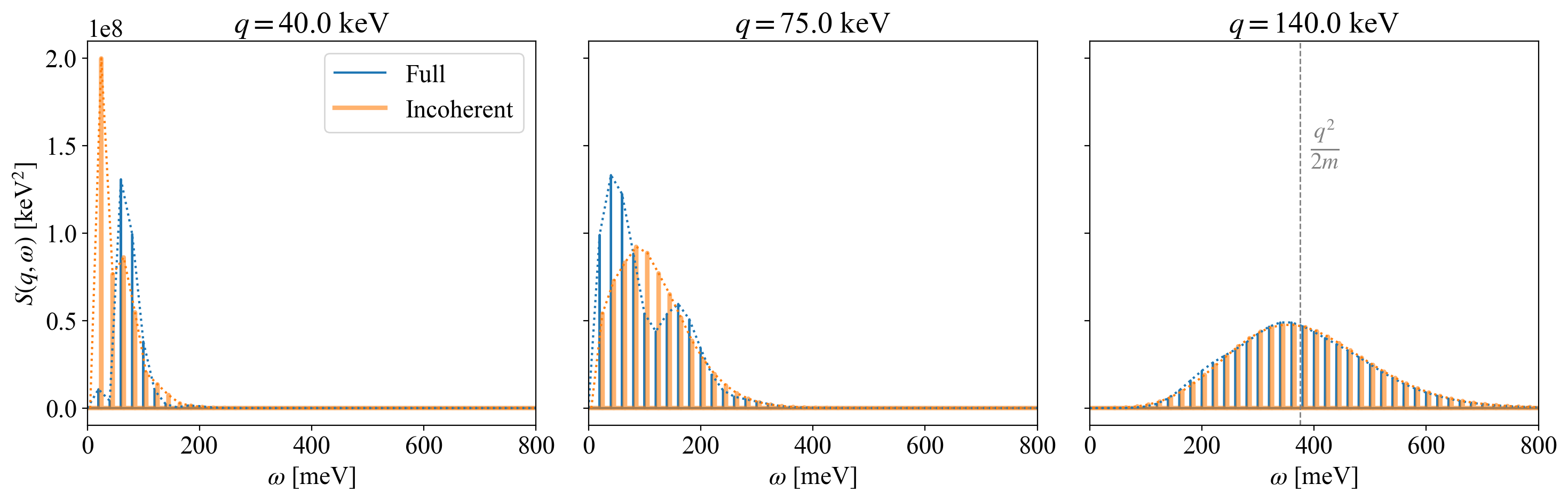}
    \caption{The structure factor $S(q,\omega)$ as a function of energy transfer $\omega$, evaluated at fixed momentum transfers of 40, 75, and 140 keV. The results are summed over all $n$. Because the exact structure factor consists of discrete $\delta$-functions, a coarse binning method (described in Eqn. \eqref{eqn:binning}) is applied to display the peaks with finite amplitudes. A dotted envelope connects these peaks to better illustrate the overall trend. At high $q$, this envelope approaches a Gaussian shape centered around the nuclear recoil limit. For visibility, the incoherent approximation is artificially shifted to the right by 5 meV. }\label{fig:2state}
\end{figure*}

We next consider the structure factor as a function of $\omega$ in Fig. \ref{fig:2state}, assuming fixed $q=40,75$, and $140$ keV.  Here, we display the sum over all phonon orders $n$ rather than plotting them individually.
To visualize the delta functions, we apply a coarse binning method for $S(q,\omega)$. For structure factor composed of delta functions as $S(q,\omega)=\sum_iS(q,\omega=\omega_i)\delta(\omega-\omega_i)$, we integrate over each bin $k$ and divide by bin size $\Delta \omega$ for normalization:
\begin{equation}
    S_\text{binned}(q,\omega_\text{$k$}) =\frac{1}{\Delta \omega} \sum_{\omega_i \text{ in $k$th bin}}S(q, \omega=\omega_i)\label{eqn:binning}
\end{equation}
The bin size used in the graph is $0.325$ meV. %

Fig. \ref{fig:2state} shows the convergence of the structure factor toward the incoherent approximation at high $q \gtrsim \sqrt{2 m \bar \omega} \approx 46$ keV, where many phonons start to be produced.
As discussed previously, the oscillating coherent corrections average to zero when summed over many terms due to the $(-1)^{n-j}$ phase factor. In this case, the sum is over many terms with different number $n$ of phonons. The full structure factor approaches a Gaussian shape centered at the nuclear recoil energy $\frac{q^2}{2m}$, which is the behavior expected from the impulse approximation, Eqn.~\eqref{eq:S_IA}.

We note that the values $\omega_+ = 20$~meV and $\omega_- = 60$~meV are chosen such that $\omega_-$ is an integer multiple of $\omega_+$. This choice is made purely for visualization. When $\omega_- = 3\omega_+$, every accessible final-state energy $\omega = j\omega_+ + (n-j)\omega_-$ collapses onto an integer multiple of $\omega_+$, so the spectrum consists of sharp lines spaced by $\omega_+ = 20$~meV. More importantly, distinct processes with different phonon numbers $(n,j)$ become exactly degenerate at each such energy. These degenerate contributions carry the alternating phase $(-1)^{n-j}$, so their $\cos(qa)$ interference terms partially cancel within a single energy, and the full structure factor agrees with the incoherent approximation bin-by-bin in the right panel of Fig. \ref{fig:2state}. Had we instead chosen incommensurate frequencies, the delta functions would be nondegenerate and would densely populate the energy axis, each retaining its individual $\cos(qa)$ correction; the cancellation would then appear only after coarse-graining over many neighboring lines. 

We summarize the findings of the toy model as follows: 
\begin{itemize}
        \item Interference terms lead to oscillatory $\cos(qa)$ behavior within individual terms of the structure factor $S(q,\omega)$.  
        \item The difference in the incoherent and full structure factors diminishes with larger number of phonons produced. As phonon order $n$ increases, an alternating phase factor in the $\cos(qa)$ coefficients of different final states leads to cancellations when summing over all states.
        \item When integrating over a wide range of $q \gg \pi/a$ or $\omega$ above the single phonon energy, the effect of the oscillatory terms average out, so the incoherent approximation provides an accurate method for rate calculations. 
\end{itemize}

\section{1D N-site Lattice}
\label{sec:Nsite}

We now extend our analysis to a 1D $N$-site periodic lattice. This system has many of the features of realistic 3D crystals, in particular containing discrete translation invariance and acoustic phonon modes, while being analytically tractable.

\subsection{Spectrum of states}\label{subsec:spectrum-of-states}

\begin{figure}[h]
    \centering
    \includegraphics[width=0.65\linewidth]{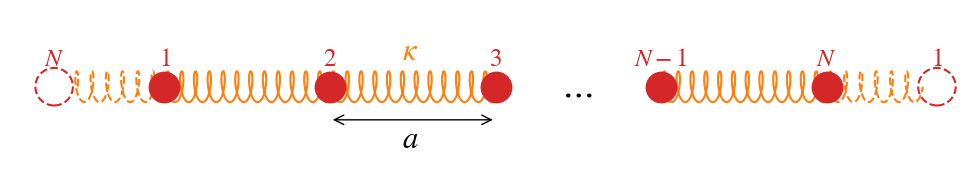}
    \caption{Schematic diagram of the 1D N-site lattice model. The system consists of $N$ identical ions of mass $m$, each connected to its nearest neighbors by springs with stiffness $\kappa$. Periodic boundary conditions are enforced by connecting the first ion directly to the $N$-th ion with an identical spring. }\label{fig:nsite-schematic}
\end{figure}

Consider $N$ identical ions of mass $m$, each connected to its nearest neighbor by a spring with constant $\kappa$. We apply periodic boundary conditions by connecting the first ion to the last, as shown in Fig. \ref{fig:nsite-schematic}; such boundary effects become negligible for large $N$. In the two-site toy model, each ion was confined by a local harmonic potential, but we omit this local confinement term here to preserve the translational symmetry of the lattice. 

The equation of motion for this system is given by
\begin{equation}
    \ddot{\vec{x}} = -\begin{pmatrix}
        \frac{2\kappa}{m} & -\frac{\kappa}{m} & 0 & \dots & 0 &  -\frac{\kappa}{m} \\
        -\frac{\kappa}{m} & \frac{2\kappa}{m} & -\frac{\kappa}{m}  & 0 & \dots \\
        \vdots\\
        -\frac{\kappa}{m} &  0 & \dots  & 0 & -\frac{\kappa}{m} & \frac{2\kappa}{m}
    \end{pmatrix} \cdot
    \vec{x},
\end{equation}
from which we obtain the eigenvectors
\begin{equation}
    \textbf{e}_{\nu, \ell} = \frac{1}{\sqrt{N}}\exp\left(i\frac{2\pi \nu \ell}{N}\right), \quad \ell = 1:N, \quad \nu = 1:N-1 \label{eqn:NlatticeEigenVec}
\end{equation}
and their corresponding eigenvalues
\begin{equation}
    \omega^2_\nu =2\left(1-\cos\frac{2\pi\nu}{N}\right)\omega_0^2\equiv 2\left(1-\cos (k_\nu a)\right)\omega_0^2 \label{eqn:NlatticeEigenVal}
\end{equation}
Or equivalently, 
\begin{equation}
    \omega_\nu = 2\omega_0\sin \frac{k_\nu a}{2}
\end{equation}
where $\omega_0^2\equiv \kappa/m$, $k_\nu \equiv \frac{2\pi\nu}{Na}$. Eqn. \eqref{eqn:NlatticeEigenVal} is the phonon dispersion relation of the $N$ site lattice system. For any $N$, there exists a zero-energy mode corresponding to a uniform translation of the entire lattice. To exclude this trivial mode, we restrict the mode index $\nu$ to $1\leq \nu\leq N-1$. 

To set a numerical value for $\omega_0$, we consider the long-wavelength continuum limit ($k_\nu\to 0$ as $N\to\infty$). In this regime, the phonon energy of the lattice reduces to a linear dependence on momentum, $\omega_\nu = k_\nu a\omega_0$.
We can map this directly to the macroscopic acoustic dispersion relation of silicon, $\omega= c_l k$, and use the speed of sound for longitudinal waves in silicon ($c_l=8433$ m/s) to set $\omega_0 = c_l/a \approx 10 \text{ meV}.$%

\begin{figure}
    \centering
    \begin{tabular}{ccc}
    \includegraphics[width=0.307\textwidth]{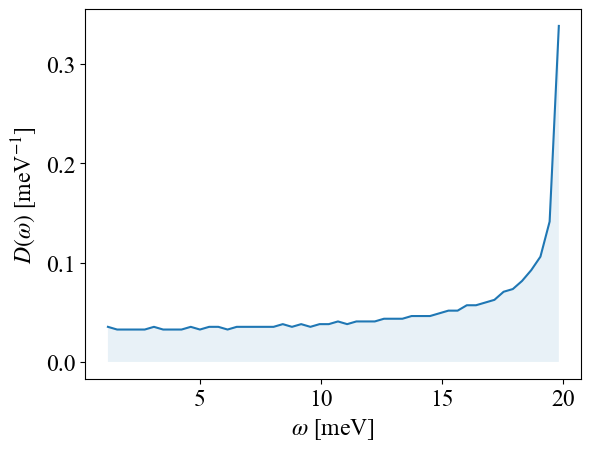} & 
    \includegraphics[width=0.32\textwidth]{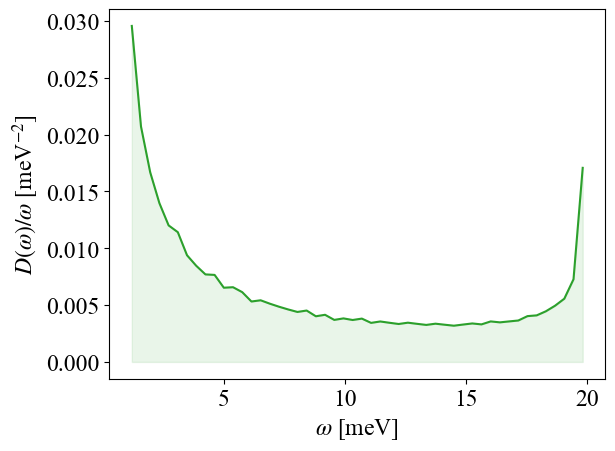} &
    \includegraphics[width=.3\textwidth]{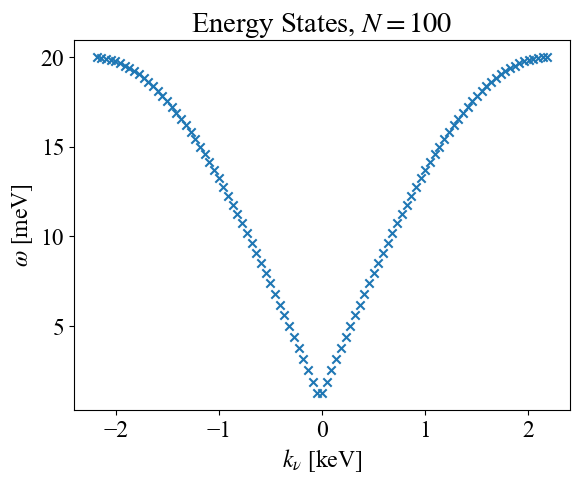}\\
        (a) & (b) & (c)
    \end{tabular}
    \caption{Phonon properties of the 1D N-site lattice. (a) The single-phonon density of states $D(\omega)$, evaluated in the large-N limit ($N=2000$) and normalized such that $\int d\omega\,D(\omega)=1$. A Van Hove singularity causes $D(\omega)$ to diverge near the maximum phonon energy ($\approx$ 20 meV) where the dispersion band becomes flat.
    (b) The ratio $D(\omega)/\omega$, which appears in the incoherent dynamical structure factor. The $1/\omega$ scaling gives an infrared divergence as $\omega \rightarrow 0$, which we regularize by a 1 meV cutoff. 
    (c) The single-phonon dispersion relation for $N=50$, with the momentum range from $-q_{BZ}$ to $q_{BZ}$. } \label{fig:nsite-dos}
\end{figure}

Fig. \ref{fig:nsite-dos}(a) illustrates the single-phonon density of states (DOS), $D(\omega)$, for large $N$. The DOS is normalized such that $\int d\omega\,D(\omega)=1$. We note that the $D(\omega)$ is finite in the limit $\omega \to 0$, which is a property of the 1D system we study: in a standard 3D crystal lattice, the DOS vanishes as $\omega \to 0$ due to the Jacobian for 3D momenta, which gives  $D(\omega) \propto k^2 \propto \omega^2$. This will lead to some differences with scattering in a 3D system, as we will see below. Meanwhile, as $\omega$ increases, $D(\omega)$ remains relatively constant before increasing drastically as it approaches the maximum phonon energy. The divergence near maximum phonon energy is due to the Van Hove singularity, where the phonon dispersion relation becomes flat, as shown in Fig. \ref{fig:nsite-dos}(c).

As shown in Eqn. \eqref{eqn:s-dos}, the structure factor (in the incoherent approximation) depends on $D(\omega)/\omega$. This quantity is plotted in Fig. \ref{fig:nsite-dos}(b). The $1/\omega$ scaling in $D(\omega)$ leads to an infrared (IR) divergence as $\omega\to 0$, which is again specific to the 1D case. In our calculations, this divergence will be regularized by a low-frequency cutoff, $\omega_\text{cut} = 1$ meV, which represents the detector energy threshold. Extremely soft phonon modes with energies below this threshold are discarded in subsequent calculations. The peaks at $\omega\to$ 0 and $\to$ 20 meV indicate that the spectral weight of the structure factor is dominated by contributions from the lowest and highest energy states, whereas in a 3D lattice $D(\omega)/\omega$ is dominated primarily by the higher energy phonons. While this difference in $D(\omega)/\omega$ does modify differential rate spectra relative to the 3D case, our main goal is use the 1D model to elucidate the role of interference effects and validate approximation schemes, which we can do regardless of the difference in the density of states.

\subsection{Structure Factor} \label{sec:nsite-sfactor}

Utilizing the normal modes derived above, we now construct the analytical form of the dynamical structure factor. The displacement of the $\ell$-th particle is given by 
\begin{equation}
    u_\ell(t) = \sum_\nu \frac{1}{\sqrt{2mN\omega_\nu}} (e^{i2\pi\nu\ell/N}  a_\nu e^{-i\omega_\nu t} +e^{-i2\pi\nu\ell/N} a_\nu^\dag e^{i\omega_\nu t})
    \label{eq:Nsite_u}
\end{equation}
which allows us to calculate the Debye-Waller factor via Eq. \eqref{eqn:DW-def}:
\begin{equation}\label{eqn:exactDW}
    W(q) =\frac{1}{2} \langle (qu_\ell(0))^2 \rangle = \frac{q^2}{4mN} \sum_{\nu=1}^{N-1} \frac{1}{\omega_\nu} = \frac{q^2}{4mN}\sum_\nu \frac{1}{2\omega_0 \sin\frac{\pi \nu}{N}}
\end{equation}
Note that in the large $N$ limit, $W(q)$ has a log-$N$ divergence because we are working in 1D; this is the same divergence discussed earlier. Our 1 meV cutoff on the energy eigenvalues removes this divergence.

Inserting the expansion Eq.~\eqref{eq:Nsite_u} into Eq.~\eqref{eqn:cll-def} yields the auto-correlation functions for the same site ($C_{\ell\ell}$ when $\ell=\ell'$) and distinct sites ($C_{\ell\ell'}$ when $\ell\neq\ell'$): 
\begin{align}
    C_{\ell\ell} &= \frac{2\pi }{V} e^{-2W(q)} \sum_{n=0}^\infty \frac{1}{n!} \left(\frac{q^2}{2mN}\right)^n \prod_{i=1}^{n} \left[  \sum_{\nu_i} \frac{1 }{\omega_{\nu_i} } \right] \ \delta {\Big (} \omega-\sum_i  \omega_{\nu_i} {\Big )} \label{eqn:NSiteCDiag}\\
    C_{\ell\ell'} &=  \frac{2\pi}{V} e^{-2W(q)}e^{iqa(\ell - \ell')} \sum_{n=0}^\infty \frac{1}{n!} \left(\frac{q^2}{2mN}\right)^n \prod_{i=1}^{n} \left[ \sum_{\nu_i} \frac{e^{-i 2\pi\nu_i(\ell-\ell')/N} }{\omega_{\nu_i} } \right] \ \delta {\Big (}\omega-\sum_i  \omega_{\nu_i} {\Big )}\\
    &=  \frac{2\pi }{V} e^{-2W(q)} \sum_{n=0}^\infty \frac{1}{n!} \left(\frac{q^2}{2mN}\right)^n \prod_{i=1}^{n}  \left[ \sum_{\nu_i} \frac{1}{\omega_{\nu_i} } \right] e^{i\left[qa-2\pi \frac{\left(\sum_{i} \nu_i \right)}{N}\right](\ell - \ell')} \ \delta {\Big (} \omega-\sum_i  \omega_{\nu_i} {\Big )}
    \label{eqn:NSiteCoffDiag}
\end{align}
Here $\nu_i$ labels the mode of the $i$th phonon in the final state.
The structure factor is then obtained by summing over these correlation functions:
\begin{equation}
    S(q, \omega) =  f^2\left[ \sum_\ell C_{\ell\ell}+ \sum_{\ell\neq\ell'}C_{\ell\ell'}\right].
\end{equation}

The key role of the off-diagonal terms is to enforce conservation of crystal momentum. To demonstrate this, we first show this explicitly for the first-order $n=1$ structure factor: 
\begin{align}
    S^{(1)}_\text{full}(q, \omega) &= \frac{2\pi f^2}{V} e^{-2W(q)} \frac{q^2}{2mN} \left[ \sum_\ell \sum_\nu \frac{\delta(\omega-\omega_\nu)}{\omega_\nu} +\sum_{\ell\neq\ell'}\sum_\nu e^{i\left[qa-2\pi\nu/N\right](\ell - \ell')} \frac{\delta(\omega-\omega_\nu)}{\omega_\nu}\right]\\
    &=  \frac{2\pi f^2}{V}e^{-2W(q)}\frac{q^2}{2mN}\sum_\nu \left[\frac{1-\cos\left(N\left(qa-\frac{2\pi\nu}{N}\right)\right)}{1-\cos\left(qa-\frac{2\pi\nu}{N}\right)}\right] \frac{\delta(\omega-\omega_\nu)}{\omega_\nu} \label{eqn:1stOrderS}  \\
    & \xrightarrow[N\to\infty]{} \frac{2\pi f^2}{\Omega_c}e^{-2W(q)}\frac{q^2}{2m}\sum_\nu \ \sum_G \delta_{q, k_\nu + G} \ \frac{\delta(\omega-\omega_\nu)}{\omega_\nu}
    \label{eqn:coherent_1phonon}
\end{align} %
In the $N\to\infty$ limit, the interference term in the square brackets in Eqn. \eqref{eqn:1stOrderS} approaches a series of delta functions, scaling as $N^2$ when $qa-2\pi\nu/N$ is an integer multiple of $2\pi$, and vanishing otherwise. This behavior enforces the condition
\begin{equation}
    q=\frac{2\pi\nu}{Na}+ \frac{2\pi\tau}{a}=k_\nu+G, \quad \tau = 0, \pm1, \pm2 \dots
\end{equation}
which yields conservation of crystal momentum up to a reciprocal lattice vector $G = 2\pi \tau /a$.
Applying this momentum constraint collapses the sum over the discrete states $\nu$, yielding the continuum limit of the single-phonon structure factor:
\begin{equation}
S_{SPC}(q, \omega) = \frac{2\pi f^2}{\Omega_c} e^{-2W(q)} \frac{q^2}{2m\omega_q} \delta(\omega-\omega_q)
\label{eq:SPC}
\end{equation}
We define this as the Single-Phonon Continuum (SPC) limit. The $\delta(\omega-\omega_q)$ term enforces momentum and energy conservation, confining the allowed scattering phase space to the single-phonon dispersion curve. Of all
the contributions to the structure factor, the single-phonon (first-order)
term is the most sensitive to the finite lattice size: at finite $N$ the sum
over $\nu$ yields discrete peaks that approach this $\delta$-function only as
$N\to\infty$. We therefore evaluate the single-phonon term in the SPC limit
rather than from a numerical $N$-site sum.

In contrast, the incoherent approximation yields a single phonon term:
\begin{align}
    S^{(1)}_\text{inc}(q, \omega) &= f^2\sum_{\ell} C_{\ell\ell} = \frac{2\pi f^2}{\Omega_c}e^{-2W(q)}\frac{q^2}{2m}\sum_\nu \frac{1}{N} \frac{\delta(\omega-\omega_\nu)}{\omega_\nu} \label{eq:incoherent_1phonon} \\
    & = \frac{2\pi f^2}{\Omega_c}e^{-2W(q)}\frac{q^2}{2m} \frac{D(\omega)}{\omega}
\end{align}
This result has the same form as Eqn. \eqref{eqn:s-dos}, where we defined the density of states:
\begin{align}
    D(\omega) &\equiv\frac{1}{N}\sum_\nu\sum_\ell|\textbf{e}_{\nu,\ell}|^2\delta(\omega-\omega_\nu) \\
    &= \frac{1}{N}\sum_\nu \delta(\omega-\omega_\nu).
\end{align}
We see that the incoherent structure factor exactly parallels the structure of the full calculation Eqns.~\eqref{eqn:1stOrderS}-\ref{eqn:coherent_1phonon}, but without the additional enforcement of momentum conservation. The incoherent structure factor thus smears the response over all phonons.

The appearance of momentum conservation in the full structure factor holds more generally for the $n$-phonon term: 
\begin{align}
\boxed{
\begin{aligned}
    S^{(n)}_{\rm full}(q, \omega) &= \frac{2\pi f^2}{\Omega_c}  \frac{e^{-2W(q)}}{n!} \left(\frac{q^2}{2m}\right)^n \prod_{i=1}^{n}  \left[ \frac{1}{N} \sum_{\nu_i} \frac{1 }{\omega_{\nu_i} } \right] \ \delta {\Big (} \omega-\sum_i  \omega_{\nu_i} {\Big )} \times \sum_G N \, \delta_{q, \sum_i k_{\nu_i} + G} \\
    S^{(n)}_{\rm inc}(q, \omega) &= \frac{2\pi f^2}{\Omega_c} \frac{e^{-2W(q)}}{n!} \left(\frac{q^2}{2m}\right)^n \prod_{i=1}^{n}  \left[ \frac{1}{N} \sum_{\nu_i} \frac{1 }{\omega_{\nu_i} } \right] \ \delta{\Big (}\omega-\sum_i  \omega_{\nu_i} {\Big )} 
\end{aligned}
}
\label{eq:Sn_key_comparison}
\end{align}
This result makes it transparent that the difference in the two structure factors is entirely the appearance of the momentum-conserving delta function. This also gives intuition for why the interference terms should be less important  when more phonons are produced. For multiple phonons produced, momentum conservation is a weak constraint on possible combinations of phonons that can appear in the final state. Conditioning on total momentum barely affects the spectrum of phonons that can appear and thus averaging the response over the density of states becomes a good approximation. In contrast, for a single phonon, momentum conservation requires  the response to sit on the single phonon dispersion. In addition, when integrating over a wide range of $q$ relative to $q_{BZ}$, differences will also average out since we sweep over all phonon momenta within the Brillouin Zone.

We can further write results in terms of the phonon density of states, take the continuum limit, and generalize our result back to $d=3$. We first define a partial density of states $\mathcal{D}(\omega, \bf k)$ in both $\omega$ and $\bf k$ by splitting the sum over modes $\nu$ into branches $\alpha$ and momentum $\bf k$:
\begin{align}
    D(\omega) &= \frac{1}{3N} \sum_\alpha \sum_{\bf k} \delta(\omega - \omega_{\alpha, \bf k} ) =  \int \frac{ d^3 {\bf k} }{(2\pi)^3} \frac{\Omega_c}{3} \sum_\alpha \delta(\omega - \omega_{\alpha, \bf k} )   \\
    \mathcal{D}(\omega, \bf k) & \equiv  \frac{\Omega_c}{3 (2\pi)^3} \sum_\alpha \delta(\omega - \omega_{\alpha, \bf k} ) , 
\end{align}
where the $1/3$ comes from the fact that there are $3N$ total modes for $N$ atoms in a 3D crystal. Then, inserting factors of unity, $\int d\omega_i \, \delta( \omega_i - \omega_{\nu_i})$, the structure factor can be written as
\begin{align}
\boxed{
\begin{aligned}
    \begin{split} S^{(n)}_{\rm full}(q, \omega) &= \frac{2\pi f^2}{\Omega_c}  \frac{ e^{-2W(q)} }{n!}   \left(\frac{q^2}{2m}\right)^n \left(\prod_{i=1}^{n}\int  d^3 {\bf k}_i \, d\omega_i \, \frac{\mathcal{D}(\omega_i, {\bf k}_i)}{\omega_i}\right)\delta {\Big (}\omega - \sum_j\omega_j {\Big )}  \\
    & \hspace{4cm} \quad \quad  \quad  \quad \times \frac{ (2\pi)^3 }{\Omega_c} \sum_{\bf G} \delta^{(3)} ( \bfq - \sum_i {\bf k}_i - {\bf G} ) \end{split} \\
    S^{(n)}_{\rm inc}(q, \omega) &= \frac{2\pi f^2}{\Omega_c} \frac{ e^{-2W(q)} }{n!}   \left(\frac{q^2}{2m}\right)^n \left(\prod_{i=1}^{n}\int d\omega_i\frac{D(\omega_i)}{\omega_i}\right) \delta {\Big (}\omega - \sum_j\omega_j {\Big )}
    \end{aligned}
    }
\label{eq:Sn_key_comparison_3d}
\end{align}
This agrees with Eqn.~\eqref{eqn:s-dos} for the incoherent result, where we sum over possible combinations of phonons, weighted by $D(\omega)/\omega$ and subject to an energy conservation constraint. In the full structure factor, the sum over modes is weighted by $\mathcal{D}(\omega, \bf k)$ and includes an integration over possible momenta combined with the momentum conservation constraint.  Note that here we have still assumed a monoatomic, isotropic lattice for simplicity.

\subsection{Numerical Results} \label{sec:nsite-sfactor_numres}

\begin{figure}[t]
    \centering
   \includegraphics[width=0.85\textwidth]{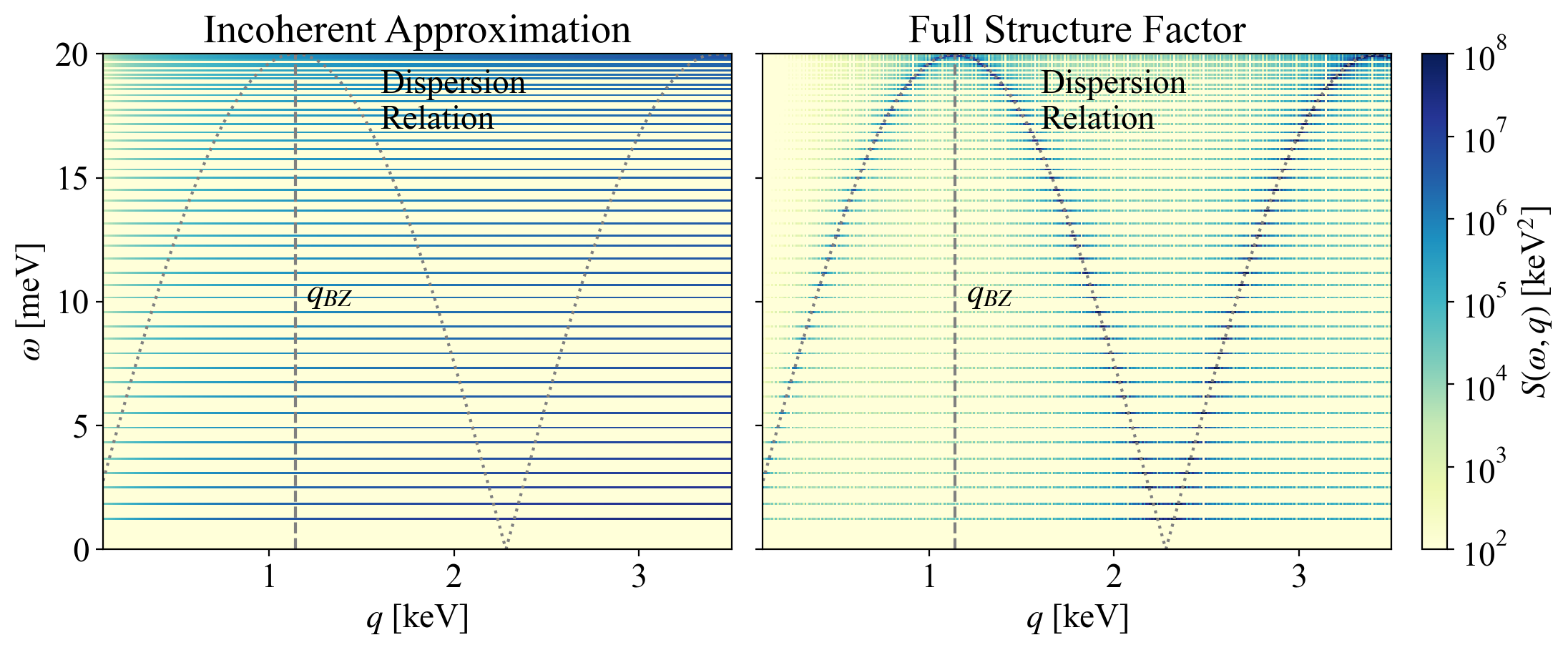}
\caption{Comparison of the structure factor in the single-phonon regime with $N = 50$. The left panel shows the incoherent approximation, in which the structure factor is broadly distributed across all momentum transfers $q$. The right panel shows the full structure factor, which exhibits a highly localized tracing along the dispersion relation $\omega_\nu = 2 \omega_0 \sin \frac{q a}{2}$.  A 1 meV cutoff is applied to both panels to address IR divergence.} \label{fig:dispersion-2compare}
\end{figure} 

\begin{figure}[t]
    \centering
        \includegraphics[width=0.9\linewidth]{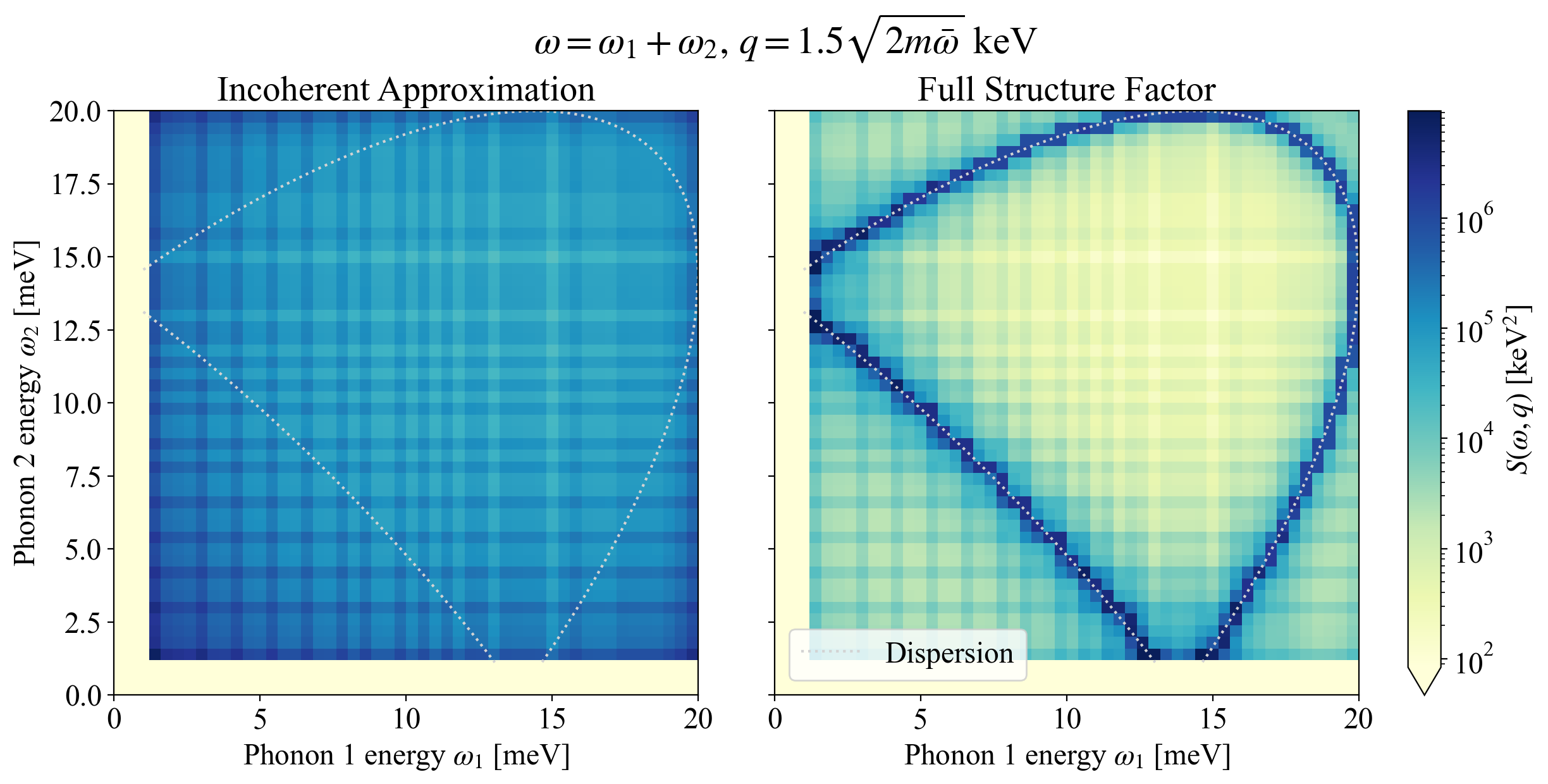}
\caption{Comparison of the two-phonon structure factor at $q=1.5\sqrt{2m\bar\omega}$ for
$N=200$, computed in the incoherent approximation (left) and from the full
structure factor (right). A coarse binning method
described in Eqn.~\eqref{eqn:binning} is applied in both panels: the value
displayed in each grid cell represents the structure factor within that grid of
width $\Delta \omega \equiv \Delta \omega_1 = \Delta\omega_2$, so the full
structure factor can be evaluated by a direct sum over all grid cells. The gray
dotted line marks the locus of $(\omega_1,\omega_2)$ allowed by the dispersion
relation with conservation of crystal momentum, $k_1 + k_2 = q + G$, enforced. The full structure factor traces
this line closely, whereas the incoherent approximation, in which the two phonon
energies are uncorrelated, spreads its weight diffusely across the kinematically
accessible region and concentrates instead at small $\omega_1$ or $\omega_2$.} \label{fig:dispersion-2ndorder}
\end{figure}

We now present the numerical calculations of the structure factor in the 1D N-site lattice. To evaluate these expressions efficiently for a large number of phonon, we used a recursive relation relating the $n$-phonon structure factor to the $n-1$ structure factor. The details and derivation of this relation are provided in Appendix \ref{app:recursive-relation}.

A comparison between the incoherent approximation and the full structure factor in the single-phonon regime is presented in Fig. \ref{fig:dispersion-2compare}. In the right panel, the full structure factor traces the dispersion relation $\omega_\nu = 2 \omega_0 \sin \frac{q a}{2}$ with conservation of momentum constraint $q = k_\nu +G$. The discrete horizontal dark and light stripes arise because the finite $N$-site lattice system has a finite number of allowed energy states. In the $N\to\infty$ limit, these discrete states will merge, rendering the energy spectrum continuous. Additionally, the non-zero width of the dispersion relation function and the non-zero off-resonance structure factor are also due to the finite $N$ value used in the calculation. 

In contrast, the dispersion relation does not appear at all in the incoherent approximation shown in the left panel. Without the momentum conservation constraint, the structure factor at each energy level is now broadly distributed across all momentum transfers $q$, rather than being localized along the dispersion curve. 

\begin{figure*}[t]
\begin{minipage}{\textwidth}
    \centering
    \includegraphics[width=0.99\linewidth]{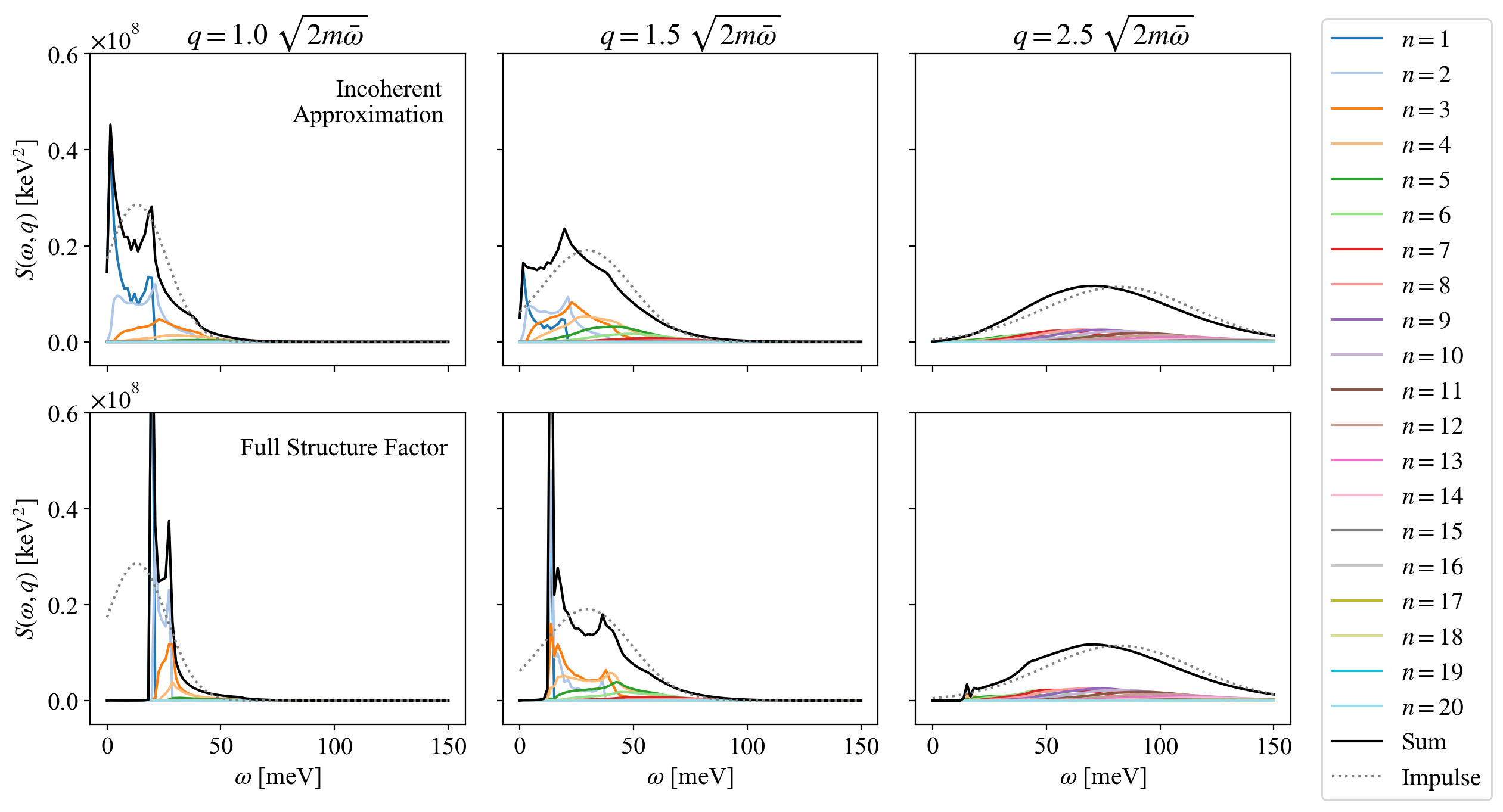}
    \caption{The comparison between the incoherent approximation (first row) and the full structure factor (second row) at different $q$ values for $N=200$. The solid black line is the sum of all orders of the structure factor, and the dotted gray line is the impulse approximation. At large $q$, the sum converges to a Gaussian envelope (the impulse approximation) for both the incoherent approximation and the full structure factor.}
\label{fig:incoh-vs-tot}
\end{minipage}
\end{figure*}

These differences can also be observed in the second-order structure factor. Fig.~\ref{fig:dispersion-2ndorder} maps the second-order structure factor as a function of the individual constituent phonon energies, $\omega_1$ and $\omega_2$ for $q=1.5\sqrt{2 m \bar\omega}$, where $\bar \omega \approx 13$ meV in this model. In the left panel, the incoherent approximation is shown, with a broad response smeared over the full spectrum of phonons for both $\omega_1$ and $\omega_2$. In the right panel, the full structure factor is shown. In this case, the allowed combination of final states is confined along the dotted contour, which represents the kinematic phase space governed by the dispersion relation and momentum conservation. In the high-intensity regions at the left and the bottom of this contour, one phonon is soft ($\approx 1$ meV with our cut), meaning the process is dominated by a single phonon absorbing the momentum. The broad, flat upper arch of the contour where one phonon has maximum energy (20 meV) and the other has approximately 15 meV represents the region where $k_1 + k_2$ is outside of the first Brillouin Zone.

Fig.~\ref{fig:incoh-vs-tot} shows the incoherent approximation and the full structure factor as a function of $\omega$ at fixed $q$, using the coarse binning method described in Eqn.~\eqref{eqn:binning}. Our values of $q$ are selected in units of $\sqrt{2 m \bar \omega}$, with $\bar \omega$ the average phonon energy, since this is the momentum scale governing the transition to many phonons produced and the nuclear recoil regime. Both the incoherent approximation and the full structure factor approach the impulse approximation, Eqn.~\eqref{eq:S_IA}, at large $q$. 

At smaller $q$, the incoherent approximation differs in the details of the target response. For $n=1$, momentum conservation dictates that only a single phonon mode on the dispersion relation can be excited, corresponding to the peak at $\omega = 2\omega_0\sin \frac{q a}{2}$ in the full structure factor. Furthermore, this dispersion relation defines the system's energy threshold, as no response can exist below it. One can see the full structure factor vanishes below the first peak, as no phonons are generated and the system is not excited, when the transferred DM energy is less than  $2\omega_0\sin \frac{q a}{2}$. The incoherent approximation smears this response over all phonons, allow for a nonzero structure factor below this threshold. %

The $n=2$ behavior of the full structure factor can also be understood in terms of the momentum-conservation constraint and correlations of the phonons in Fig.~\ref{fig:dispersion-2ndorder}. At $q=1.5\sqrt{2m\bar\omega}$, there are peaks in the $n=2$ term near $\omega = $ 15 meV and 35 meV. These are precisely the regions identified in Fig.~\ref{fig:dispersion-2ndorder}: the 15 meV peak corresponds to one phonon absorbing most of the energy and momentum, and the 35 meV peak arises because the dispersion curve is relatively flat there, such that a large number of states accumulate within this region.
Again, the incoherent approximation does not enforce the above constraints, and the response is smeared over the full spectrum of phonons. 
These detailed differences in the specific phonons produced will generally wash out when we consider integrated responses over a range of $q$ and $\omega$, however, similar to what we found in Sec.~\ref{sec:2site}. In particular, integration over a wide range of $q$ essentially removes the momentum-conservation constraint in Eqn.~\eqref{eq:Sn_key_comparison}, yielding identical results in the full and incoherent structure factor.

\subsection{Behavior of $S(q,\omega)$ across $q$ and $\omega$}

\begin{figure*}[t]
    \centering
    \includegraphics[width=0.95\textwidth]{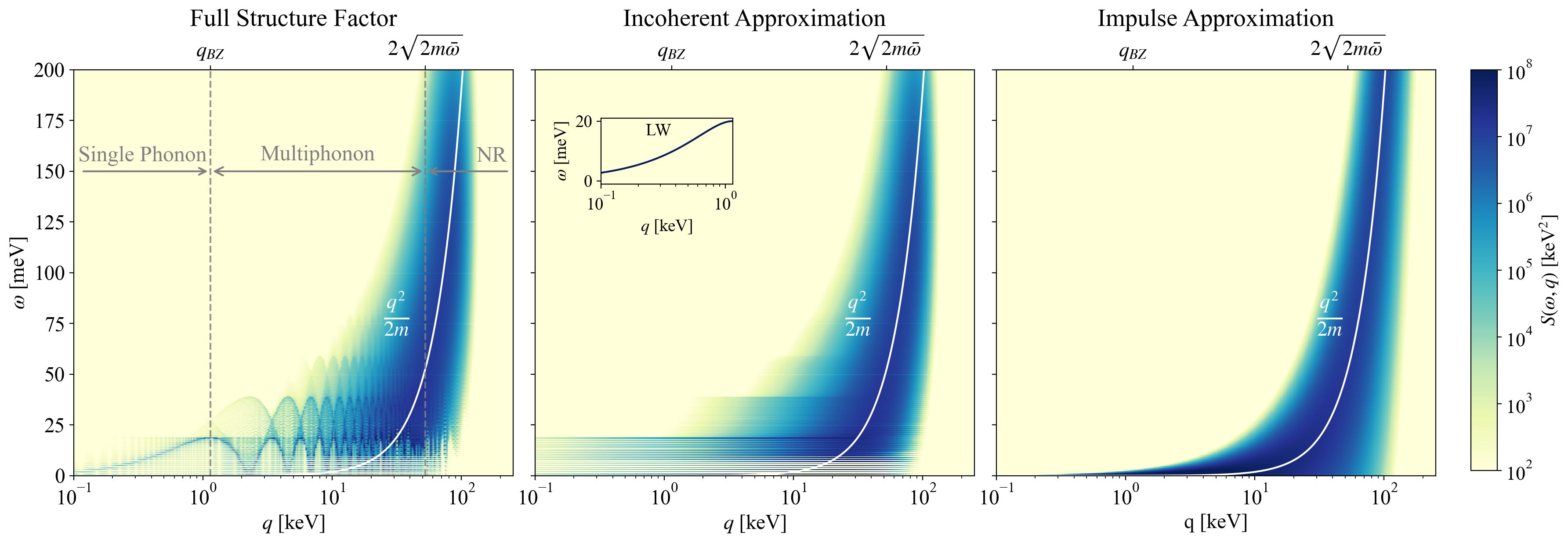}
    \caption{Comparison of the structure factor evaluated using different methods. The panels display the exact full structure factor (left), the incoherent approximation with an inset for the long-wavelength approximation (middle), and the impulse approximation (right). At low momentum transfers ($q < q_{BZ}$), the full structure factor captures the single phonon dispersion relation that the incoherent and impulse approximation miss. At high momentum transfers ($q \gg \sqrt{2 m \bar \omega}$), both the full and incoherent structure factors converge to the impulse approximation. The solid white curve in each panel denotes the free nuclear recoil limit, $\omega = q^2/2m$.} \label{fig:sfactor-3compare}
\end{figure*}

To provide an overview of its behavior over a large range of $\omega$ and $q$, we show the structure factor calculated by the full structure factor (left), incoherent approximation (middle), and  impulse approximation (right) in Fig. \ref{fig:sfactor-3compare}. 
The presence of the single phonon dispersion can be clearly seen in the full structure factor. This dispersion is inherited by higher order terms, as can be seen in structures above the single phonon energy. Another feature captured by the full structure factor is a kinematic threshold, where $S(q,\omega)$ vanishes if the energy transfer is below the lowest single-phonon energy allowed by the dispersion relation for all $q$ values.

The incoherent approximation differs notably from the full structure factor for $q < q_{BZ}$ within the single phonon regime, due to the removal of the momentum conservation constraint. For $q > q_{BZ}$, the incoherent approximation captures the broad features of the full structure factor, although it differs in the specific structures below the scale $q = \sqrt{2 m \bar \omega}$. While using an incoherent structure factor does not properly capture momentum conservation, this becomes less important as more phonons are produced.

Since the incoherent approximation gives a much faster method to compute the structure factor, particularly if one considers 3D crystals, it would be valuable to stitch it together with elements of the full structure factor at low-$q$ for fast and accurate results. Similar to Ref.~\cite{Campbell-Deem:2022fqm}, we will employ a long-wavelength approximation at $q < q_{BZ}$ for the single phonon term:
\begin{equation}
    S_{LW}\approx\frac{2\pi f^2}{\Omega_c}\frac{q^2}{2 m\omega_{q}}\delta(\omega-\omega_{q})
    \label{eq:S_LW}
\end{equation}
where the dispersion is $\omega_{q} = 2\omega_0 \sin\frac{qa}{2}$.
This is illustrated in the inset of the middle panel. The long-wavelength approximation is a simplification of the single-phonon continuum limit in Eqn.~\eqref{eq:SPC}, where in taking the long-wavelength limit ($q \to 0$), the Debye-Waller factor $\exp(-2W(q))$ approaches $1$. 
Combining this approximation with the incoherent approximation for all other $n$ and $q > q_{BZ}$ allows for rapidly calculations, while preserving the behavior of the $n=1$ full structure factor at low $q$. We will verify this numerically in the next section.

Finally, for $q \gg \sqrt{2 m \bar \omega} $, many phonons are produced and kinematic constraints from momentum conservation are completely averaged out.
In this high-momentum regime, both the full structure factor and the incoherent approximation converge to the same continuous spectrum, which can be described by the impulse approximation shown in the right panel of Fig.~\ref{fig:sfactor-3compare}.

\section{Scattering Rate}
\label{sec:scattering_rate}

We now evaluate DM scattering rates, taking our results for the 1D structure factor as  approximations for the isotropic 3D structure factor. Everything else in the setup is assumed to be the full 3D scattering rate.

In the isotropic limit, the scattering rate can be written as 
\begin{equation}
    R=\frac{1}{4\pi\rho_T}\frac{\rho_\chi}{m_\chi}\frac{\sigma_p}{\mu_\chi^2}\int_{q_-}^{q_+}dq\int^{\omega_+}_{\omega_\text{th}}d\omega \,q\,\eta(v_\text{min})|F(q)|^2 S(q,\omega),
\end{equation}
where $\rho_T$ and $\rho_\chi$ are the density of the target crystal and dark matter, respectively. $m_\chi$ is the mass of a dark matter particle, $\sigma_p$ is the DM-proton scattering cross section, and $\mu_\chi$ is the DM-proton reduced mass. 
The integration limits are given by
\begin{align}
    q_\pm &= m_\chi v_{\rm max} \left(1\pm\sqrt{1-\frac{2\omega_\text{th}}{m_\chi v_{\rm max}^2}}\right)\\
    \omega_+ &= q v_{\rm max} -\frac{q^2}{2m_\chi}
\end{align}
where $v_{\rm max} = v_{\rm esc} + v_E$ is the maximum DM speed in the lab frame.
The mean inverse dark matter speed $\eta(v_\text{min})$ is defined as 
\begin{equation}
    \eta(v_\text{min})  =\int_{v_\text{min}}^\infty \, d^3 \textbf{v} \, \frac{f_\chi (\textbf{v})}{v}
\end{equation}
where $v_\text{min} = \frac{q}{2m_\chi}+\frac{\omega}{q}$. $f_\chi(\textbf{v})$ is the Dark Matter velocity distribution, which, following the \textit{Standard Halo Model}, equals to
\begin{align}
    f_\chi & = \frac{1}{N_0} e^{-\frac{|{\bf v} + {\bf v_E}|^2}{v_0^2}}\Theta(v_{esc}-|{\bf v}+{\bf v_E}|)\\
    N_0 & = \pi^{3/2}v_0^3\left[\operatorname{erf}\left(\frac{v_\text{esc}}{v
    _0}\right) - \frac{2}{\sqrt{\pi}} \frac{v_\text{esc}}{v
    _0}e^{-\left(\frac{v_\text{esc}}{v_0}\right)^2}\right].
\end{align}
Throughout, we use $f = 28$, the mass number of Silicon.

We now compare differential scattering rate results using the full 1D structure factor with that obtained using the incoherent approximation. We also consider the combined incoherent and long-wavelength (Inc + LW) approximation, which stitches together the incoherent approximation with a single-phonon coherent result. Both the full structure factor and the incoherent approximation are
calculated using the 1D $N$-site model discussed in the previous section,
with one exception: the first-order term of the full structure factor is
evaluated in the SPC limit of Eq.~\eqref{eq:SPC} rather than from the finite
$N$-site sum. For the combined Inc + LW approximation, we utilize the long-wavelength approximation within the single phonon regime ($q<q_{BZ}$) for the single phonon structure factor, and revert to the incoherent approximation for the multiphonon regime ($q_{BZ}<q<2\sqrt{2m\bar\omega}$) for the higher order structure factor. Across all three models, we utilize the impulse approximation in the high momentum transfer regime, when $q>2\sqrt{2m\bar\omega}$.  Table \ref{table:1} summarizes the different methods used for the combined Inc + LW approximation across all momentum transfer regimes.

\begin{table}[htpb]
\centering
\begin{tabular}{l|l}
\toprule
\textbf{Condition} & \textbf{Inc + LW} \\ 
\midrule
$q < q_{BZ}$ & \makecell[l]{order $n=1$: LW approx, Eqn.~\eqref{eq:S_LW}\\ $n>1$: Incoherent, $N$-site} \\ 
\addlinespace
$q_{BZ} < q < 2\sqrt{2m\bar\omega}$ & Incoherent, $N$-site \\ 
\addlinespace
$q > 2\sqrt{2m\bar\omega}$ & Impulse approximation, Eqn.~\eqref{eq:S_IA} \\ 
\bottomrule
\end{tabular}
\caption{Summary of the combined Incoherent plus long-wavelength (Inc + LW) approximation scheme. This approach combines the much faster calculations in the incoherent approximation with the more accurate coherent single-phonon structure factor at low $q$. }
\label{table:1}
\end{table}

\subsection{$dR/dq$ for Heavy Mediator}

\begin{figure}[h]
    \centering
    \includegraphics[width=0.99\linewidth]{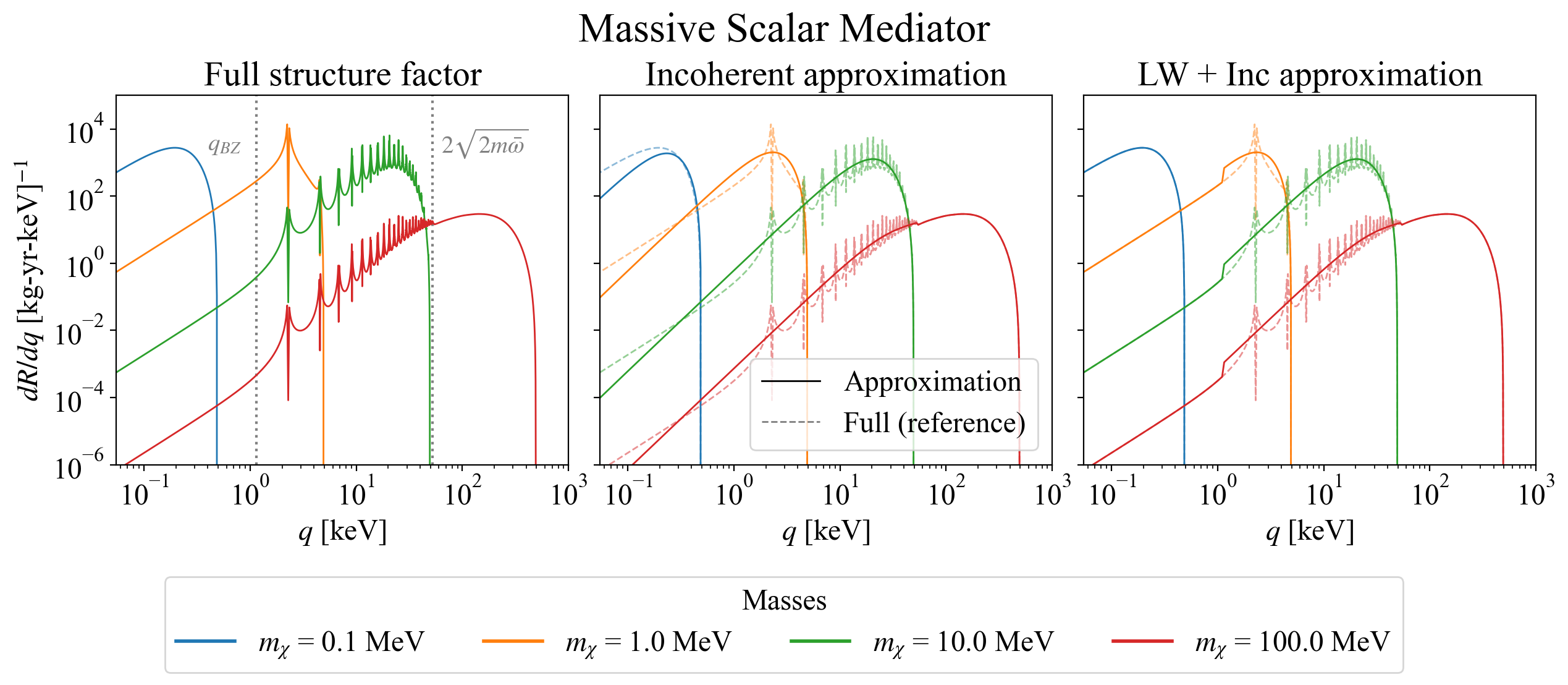}
    \caption{Differential scattering rate $dR/dq$ versus momentum transfer $q$ for a massive scalar mediator computed at $N=100$. The panels are the full structure factor (left), the incoherent approximation (center), and the combined long-wavelength plus incoherent approximation (right). In the center and right panels, the incoherent and Inc + LW approximations (solid lines) are plotted against the full structure factor (dashed lines) to demonstrate the accuracy of these two approximations. Vertical dotted lines in the leftmost panel denote the first Brillouin zone ($q_{BZ}$) and the kinematic scale ($2\sqrt{2m\bar{\omega}}$). }
    \label{fig:dRdq}
\end{figure}

To gain deeper insight into the impact of the different approximations, we analyze the differential scattering rate with respect to momentum transfer, $dR/dq$, for a massive scalar mediator (form factor ${F}(q)=1$) as presented in Fig. \ref{fig:dRdq}, calculated using $\sigma_p=10^{-40}$ cm$^2$.

First, the exact full structure factor evaluated at $N=100$ exhibits oscillatory features with sharp peaks at integer multiples of the reciprocal lattice vector $2\pi/a$. As discussed in Sec. \ref{subsec:spectrum-of-states}, these peaks arise from the IR divergence of the low-energy states, indicating that the target lattice is highly sensitive to specific momentum transfers. The sharp dips in the middle of these discrete peaks are due to the 1 meV low-frequency cutoff applied to regularize the IR divergence, without which the rate would diverge. Notably, this oscillatory pattern gradually diminishes at higher momentum transfers and ultimately vanishes around $q = 2\sqrt{2m\overline{\omega}}$. At this boundary, the scattering enters the nuclear recoil regime where all methods uniformly converge to the impulse approximation.

Second, the purely incoherent approximation exhibits significant deviations in the low-momentum regime ($q < q_{BZ}$). Because the incoherent method approximates the target as a collection of independent oscillators, it ignores the exact kinematic constraints imposed by the lattice's dispersion relation. Consequently, it predicts an incorrect power-law slope at low $q$ and produces a smooth profile that completely misses the sharp oscillatory peaks present in the full structure factor at higher $q$.

\begin{figure}
    \centering
    \includegraphics[width=0.95\linewidth]{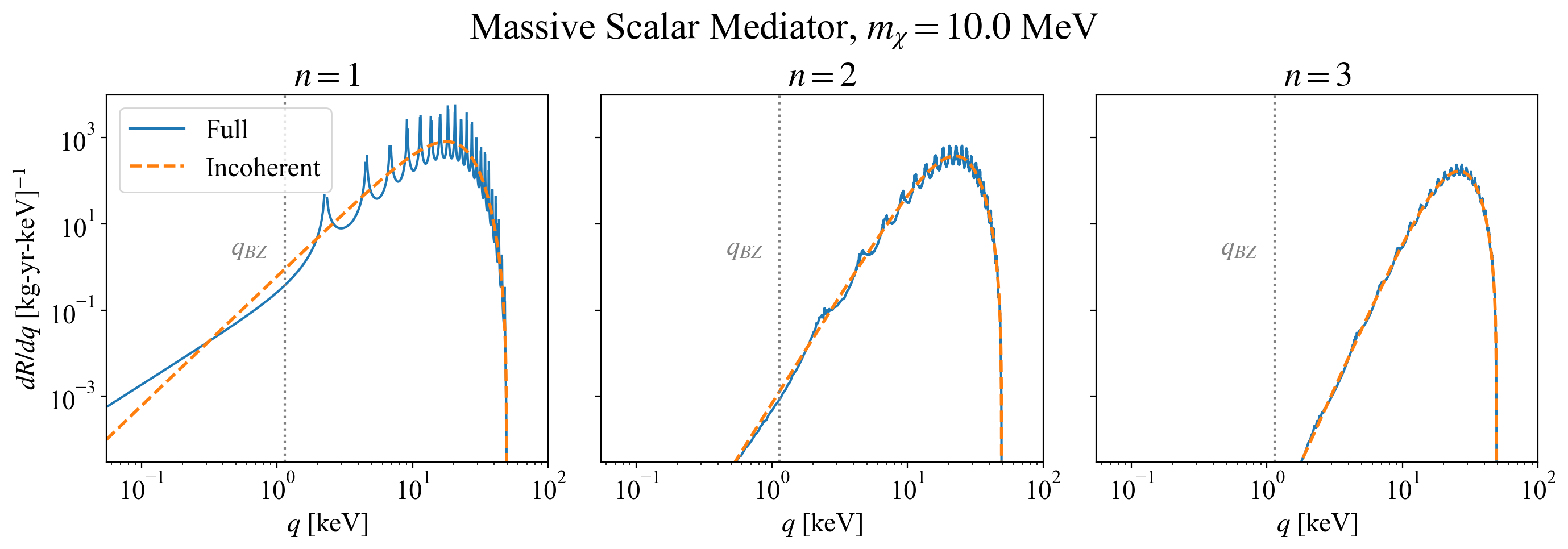}
    \caption{$dR/dq$ as a function of momentum transfer $q$ for a massive scalar mediator
with fixed dark matter mass $m_\chi = 10.0$ MeV. The three panels show the
contributions from single-phonon ($n=1$, left), two-phonon ($n=2$, center),
and three-phonon ($n=3$, right) processes, each comparing the full
calculation (solid blue) to the incoherent approximation (dashed orange).
The vertical dotted line marks the edge of the first Brillouin zone,
$q_{BZ}$. The $n=1$ panel is evaluated in the single-phonon coherent (SPC)
limit, while $n=2,3$ are computed with $N=100$. In the SPC limit the rate
vanishes at isolated values of $q$ due to the 1 meV cut we apply to the
phonon energies; these zeros are omitted from the plot to avoid the spurious vertical lines they would produce on a logarithmic axis. As the number of phonons increases from left to right, the sharp peaks smooth out and the incoherent approximation gradually matches the full structure factor. }
    \label{fig:dRdq-3orders}
\end{figure}

\begin{figure}
    \centering
    \includegraphics[width=0.99\linewidth]{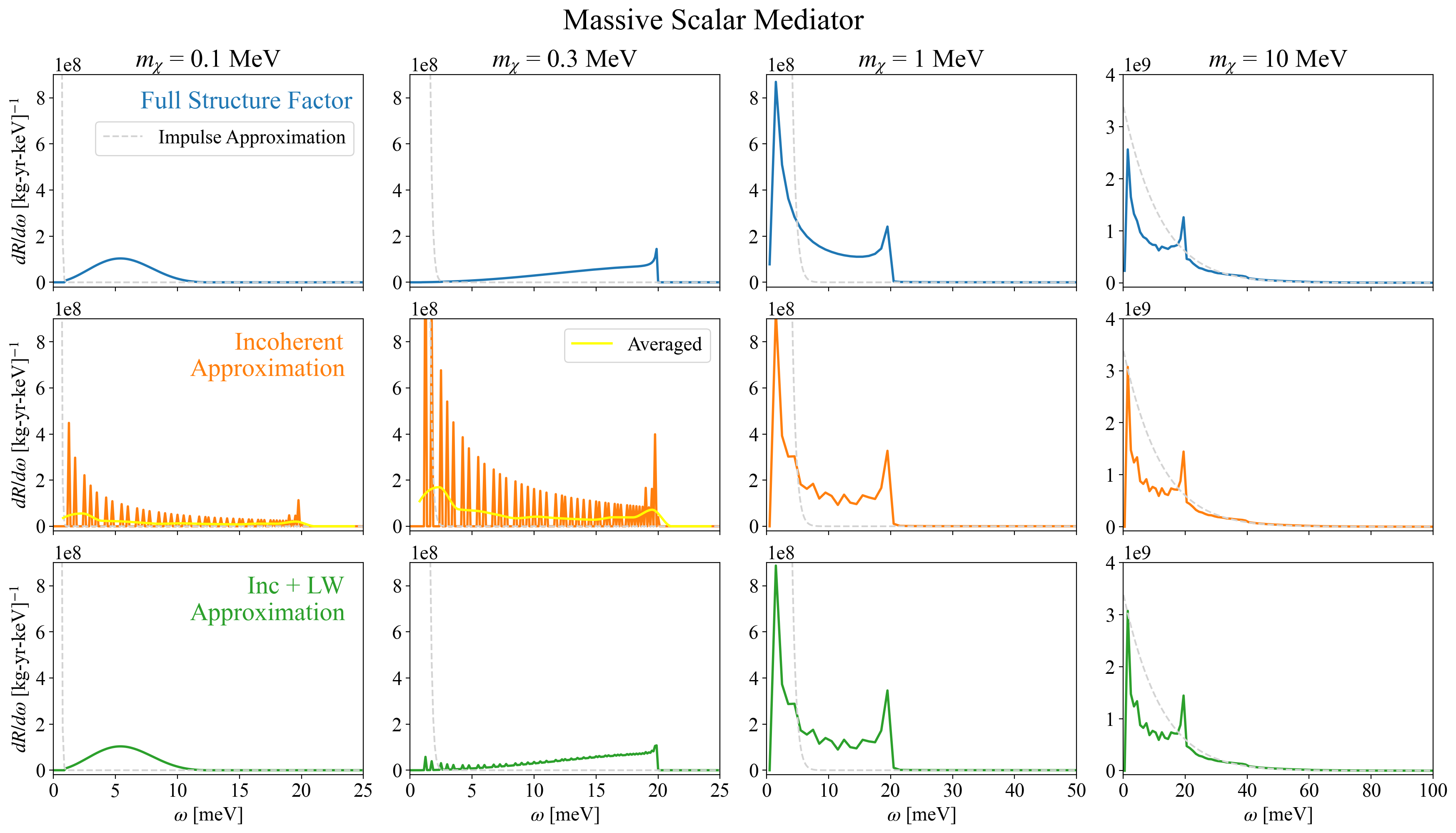}
    \caption{$dR/d\omega$ vs $\omega$ for a massive scalar mediator at different dark matter masses (columns) using three different calculation methods (rows) calculated at $N=100$. The three methods shown are the Full Structure Factor (blue), Incoherent Approximation (orange), and the combined Inc + LW approximation (green). In all panels, the dashed gray line represents the impulse approximation. Because calculating using a finite $N$ produces highly spiky spectra, we also show averaged rates in the 0.1 and 0.3 MeV columns for the incoherent approximation in yellow. In the 1 and 10 MeV columns, we omit the raw spikes entirely and plot only the averaged $dR/d\omega$ for all methods. }
    \label{fig:dRdomega}
\end{figure}

To understand why this failure is most prominent in the low-$q$ regime, it is instructive to break down the structure factor by the number of phonons ($n$), as shown in Fig. \ref{fig:dRdq-3orders}. The oscillatory pattern and the low-$q$ behavior are almost entirely driven by first-order ($n=1$) single-phonon processes. Because $n=1$ processes dominate the total scattering rate at $q < q_{BZ}$, their high sensitivity to the exact dispersion relation causes the discrepancy between the incoherent approximation and the full structure factor in this regime. Conversely, for second-order ($n=2$) and higher multiphonon processes, the size of these oscillations damp out, similar to what we observed in Fig.~\ref{fig:2state-int} for the two-site model. Therefore, at higher $q$ where multiphonon processes dominate, the lack of exact dispersion constraints becomes negligible, and the incoherent approximation becomes a faithful representation of the full structure factor.

Lastly, the combined Inc + LW approximation utilizes the long-wavelength approximation for single phonon processes at $q<q_{BZ}$ to ensure the scattering behavior for low mass dark matter is evaluated correctly. For higher $q$ where the multiphonon processes dominate, we switch to the incoherent approximation. This allows us to bypass the heavy computational cost of the exact full structure factor while still obtaining an accurate scattering rate across a broad range of dark matter masses. This approach reproduces the full structure factor exactly at small $q$. While it does not capture the detailed structures of the full structure factor at larger $q$, it describes the average behavior. Hence, any rates obtained from integrating over a wide range of $q$ will not be sensitive to these differences.

\subsection{$dR/d\omega$ for Heavy Mediator}

In addition to the $dR/dq$ distribution, we compute the $dR/d\omega$ differential rate to evaluate the energy spectrum of the rate. These results are shown in Fig. \ref{fig:dRdomega}.

The behavior of the full structure factor is strongly dictated by the kinematically allowed dark matter phase space. At $m_\chi = 0.1$ MeV, kinematic constraints from the phase space and the dispersion relation put a limit on the maximum allowed energy, causing the rate to vanish at $\omega\approx10$ meV. For $m_\chi = 0.3$ MeV, the available phase space expands to cover the full Brillouin Zone, $q \gtrsim q_{BZ}$, and the kinematically allowed energy now exceeds the maximum single-phonon energy. As a result, the rate spans from the 1 meV IR cutoff to the 20 meV single-phonon limit. In this regime, the density of states starts to influence the shape, producing a peak at the maximum phonon energy due to the Van Hove singularity. As the mass increases to $m_\chi = 1$ MeV, the phase space expands further and coherent corrections average out, resulting in a profile that closely resembles the density of states. A sharp cutoff remains at the maximum single-phonon energy, indicating that single-phonon processes still dominate. Finally, at $m_\chi = 10$ MeV, a higher energy tail indicates the transition into the multi-phonon scattering regime. 

In contrast, for $m_\chi = 0.1$ MeV, the pure incoherent approximation improperly reduces to a profile proportional to the density of states, failing to capture the kinematic cutoff indicated by the dispersion relation. Note that the incoherent approximation contains distinct spikes in the computed rates because the calculations use a finite system size of $N=100$, such that the accessible energy levels are discrete. The averaged trend is shown in yellow to guide the eye. In the full structure factor, we use the single phonon continuum, which eliminates these spikes. For heavier dark matter ($m_\chi \ge 1 \text{ MeV}$), the incoherent approximation ultimately converges to the same behavior as the full structure factor. 

Finally, the combined Inc + LW approximation successfully restores the low $q$ coherent behavior of the structure factor. It thus reproduces the behavior of the full structure factor for all DM masses shown.

\subsection{Final Result}

\begin{figure}[t]
    \centering
    \includegraphics[width=0.99\linewidth]{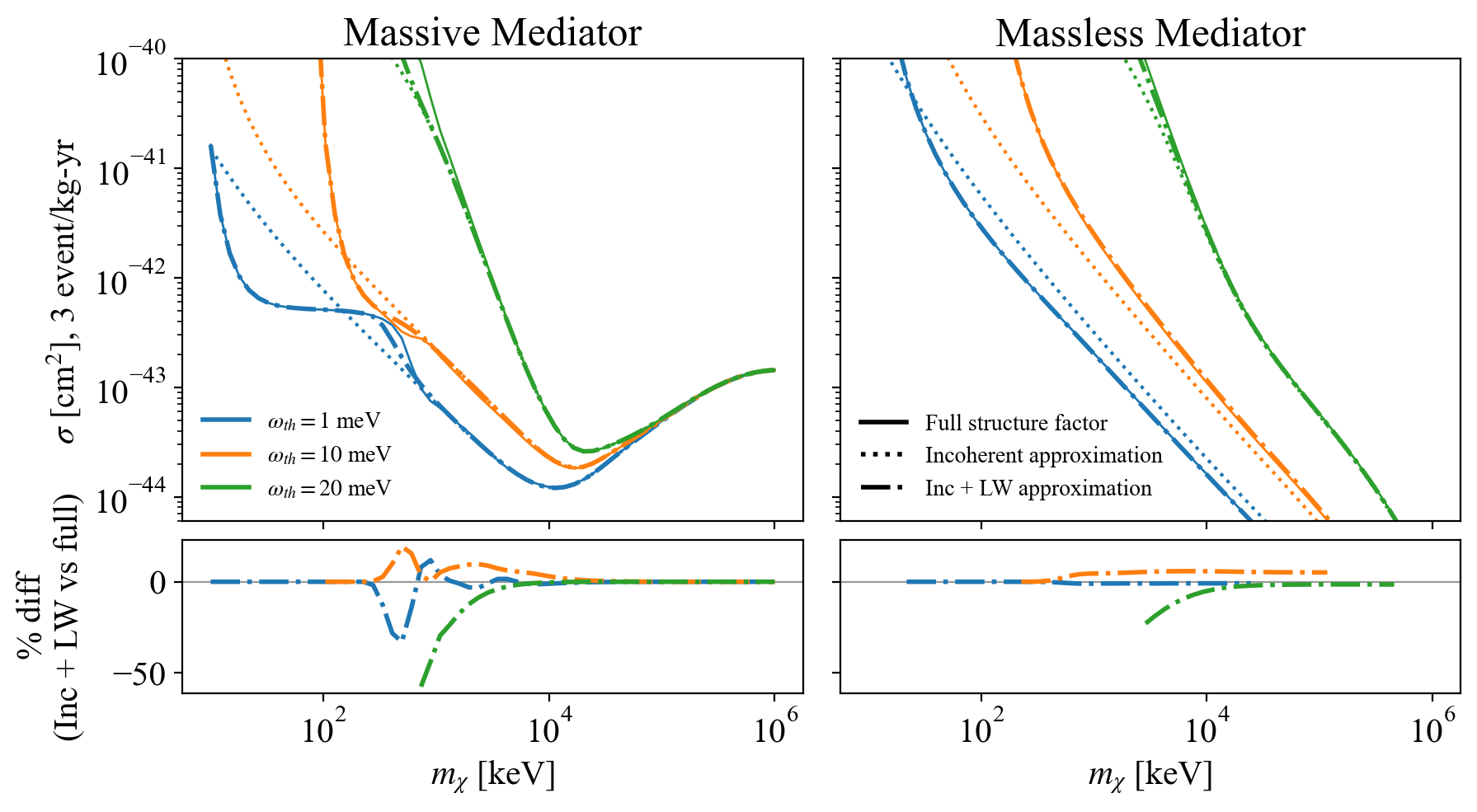}
    \caption{DM scattering cross-section $\sigma$ as a function of DM mass $m_\chi$ for massive (left) and massless (right) mediators. Different colors represent different detector energy thresholds:  $1\text{ meV}$ (blue), $10\text{ meV}$ (orange), and $20\text{ meV}$ (green). The solid lines represent the full structure factor, the dotted lines represent the incoherent approximation, and dash-dotted lines represent the combined incoherent plus long-wavelength approximation, whose fractional deviation from the full result is shown in the lower panels. }
    \label{fig:crossSection}
\end{figure}

In Fig. \ref{fig:crossSection}, we compare the impact of different structure factor calculations on the cross sections needed to achieve 3 events/kg-year. We consider detector threshold energies of $\omega_{th}=1, 10$ and $20$ meV, considering both the massive (left panel) and massless (right panel) scalar mediator cases. For the massive scalar mediator, the form factor is $\tilde{F}(q)=1$, and for the massless scalar mediator, the form factor is $\tilde{F}(q)=(m_\chi v_0/q)^2$ with $v_0=220$ km/s. Because the scattering rate scales as $1/q^4$, the kinematics strongly favor forward scattering, meaning the rate across all DM masses is heavily influenced by the low-q behavior of the structure factor.

At low detector thresholds of 1-10 meV, the low-$\omega$, low-$q$ regime is kinematically accessible. For the massive mediator, the incoherent approximation diverges from the full structure factor for $m_\chi \lesssim 1$ MeV. This breakdown is expected, since single-phonon scattering dominates at these masses, while the combined Inc + LW approximation successfully reproduces behavior of the full structure factor. For the massless mediator, the effects of this low $\omega, q$ regime are even more pronounced. %

When the detection threshold equals or exceeds the maximum single-phonon energy ($\omega_{th}=20$ meV), the scattering kinematics are driven into the multi-phonon regime. As seen in both the massive and massless scalar mediator panels, the incoherent approximation and the combined Inc + LW approximation agree results using the full structure factor, particularly at large DM mass. 

In conclusion, for both massive and massless scalar mediators, the combined Inc + LW approximation demonstrates excellent agreement with the full structure factor, robustly reproducing the predicted scattering rates to within 30\% across all evaluated masses, with the exception of masses close to the kinematic threshold for a given $\omega_{\rm th}$. %

\section{Conclusions}
\label{sec:conclusions}

The evaluation of sub-GeV dark matter scattering rates relies on the understanding of the crystal target's dynamic structure factor. In this paper, we used a 1D N-site crystal lattice model to analytically study the full structure factor and clarify the importance of interference effects in DM scattering off of crystals. Using it as a stand-in for the isotropic 3D structure factor, we can  numerically estimate the error of neglecting such interference terms in more complicated calculations with realistic phonons in 3D targets.

Our key findings are summarized as follows:
\begin{itemize}
    \item Momentum conservation from interference terms: The only difference between the full structure factor and the incoherent approximation is that crystal momentum conservation is enforced in the full structure factor, Eqns.~\eqref{eq:Sn_key_comparison} and ~\eqref{eq:Sn_key_comparison_3d}.   In the incoherent structure factor, the response is smeared over the entire phonon density of states with only energy conservation as the constraint. The transition from coherent to incoherent scattering occurs because crystal momentum conservation becomes a weak constraint on individual phonon energies and momenta when many phonons are produced. 
    
    \item Success of the incoherent approximation for multiphonon processes: The incoherent approximation, in which interference terms are neglected, proves to be a highly reliable model for two-phonon and higher-order processes ($n \ge 2$). Momentum conservation does place restrictions on the phase space of phonons produced at specific $q, \omega$ values. However, when integrating over a range of $q$ or $\omega$ values larger than $q_{BZ}$ or single phonon energies, these differences average out.

    \item Validation of a combined incoherent + long-wavelength (Inc + LW) approximation: As a consequence of the previous point, we can employ the incoherent approximation for multiphonons in direct direction calculations. We tested a approximation scheme combining coherent scattering into long wavelength phonons with incoherent scattering otherwise. We find that it 
    successfully reproduces the full structure factor’s scattering rates for both massive and massless scalar mediators, with only moderate errors in specific mass ranges or near kinematic thresholds.
    
\end{itemize}

Looking forward, as the theoretical and experimental focus continues to shift toward MeV-scale dark matter candidates, understanding scattering into multiphonons will be needed to accurately model detector sensitivities. While our analytical derivation was performed on a simplified 1D lattice model, the fundamental kinematic behaviors uncovered here are expected to persist in 3D targets. In the future, the insights obtained here can be applied to realistic 3D crystals to further improve theoretical predictions for upcoming direct-detection experiments.

\section*{Acknowledgments}

We thank Julian Dukes for insightful discussions that helped us identify the relationship in Eqn.~\eqref{eq:Sn_key_comparison}. LL and TL acknowledge support by US Department of Energy Office of Science under Award No. DE-SC0022104, a Harold and Suzy Ticho Endowed Fellowship, and the Defense Advanced Research Projects Agency under cooperative agreement HR0011-25-2-0035. MF was supported by an Undergraduate Summer Research Award (URSA) in the Physical Sciences Division at UCSD.

The views expressed in this paper are those of the authors
and do not reflect the official policy or position the Department of Defense or the U.S.
Government, and no official endorsement should be inferred. 

The authors used Claude Opus 4.8 and Gemini 3 Pro to assist with code development and debugging, and to proofread portions of the manuscript. AI-assisted code and suggestions were reviewed and validated by the authors, and the authors take full responsibility for the accuracy and content of the work.

\appendix

\section{Recursive Relation} \label{app:recursive-relation}

The solutions in Eqn.~\eqref{eqn:NSiteCDiag} and Eqn. \eqref{eqn:NSiteCoffDiag} can be time-consuming to compute. Therefore, in practice, we evaluate them using the recursive relations given below.

For the diagonal terms:
\begin{align}
    C_{\ell\ell, n+1 }& (q, \omega) =  \frac{e^{-2W(q)}}{V}\left(\frac{q^2}{2mN}\right)^{n+1} \int_{-\infty}^\infty   \frac{1}{(n+1)!}\textbf{}\left(\sum_{\nu} \frac{e^{i\omega_\nu t}}{\omega_\nu}\right)^{n+1} e^{-i\omega t} dt\\
    = &  \frac{e^{-2W(q)}}{V} \left(\frac{q^2}{2mN}\right)\left(\frac{q^2}{2mN}\right)^{n}\int_{-\infty}^\infty  \frac{1}{n+1} \frac{1}{n!}\left(\sum_{\nu} \frac{e^{i\omega_\nu t}}{\omega_\nu}\right)\left(\sum_{\nu'}\frac{e^{i\omega_{\nu'} t}}{\omega_{\nu'}}\right)^{n} e^{-i\omega t} dt\\
    =  & \frac{1}{n+1} \left(\frac{q^2}{2mN}\right) \sum_{\nu}\frac{C_{\ell\ell, n}(\omega-\omega_\nu, q)}{\omega_\nu} 
\end{align}
with 
\begin{equation}
    C_{\ell\ell, n=1}(q, \omega) = \frac{2\pi}{V}e^{-2W(q)} \left(\frac{q^2} {2mN}\right) \sum_\nu\frac{\delta(\omega-\omega_\nu)}{\omega_\nu}.
\end{equation}
\bigskip

For the interference terms:
\begin{align}
    &C_{\ell\ell', n}(\omega, q)  = \frac{e^{-2W(q)}}{V}  e^{iqa(\ell-\ell')} \left(\frac{q^2}{2mN}\right)^n \int_{-\infty}^\infty \frac{1}{n!}\left(\sum_{\nu=1}^{N-1} \frac{e^{-i 2\pi\nu(\ell-\ell')/N} }{\omega_\nu }e^{i\omega_\nu t}\right)^n e^{-i\omega t} dt\\
    &=\frac{2\pi}{V} e^{-2W(q)} e^{iqa(\ell-\ell')}\left(\frac{q^2}{2mN}\right)^n \sum_{j_1+j_2+\dots+j_N = n} \left[\delta\left(\omega-\sum_\nu j_\nu \omega_\nu\right)\prod_{\nu=1}^{N}\frac{e^{-i2\pi\nu j_\nu (\ell-\ell')/N} }{j_\nu!(\omega_\nu )   ^{j_\nu}   } \right]\\
     &= \frac{2\pi}{V} e^{-2W(q)}\left(\frac{q^2}{2mN}\right)^n \times \nonumber \\
     & \quad \quad \quad \sum_{j_1+j_2+\dots+j_N = n} \left[\delta\left(\omega-\sum_\nu j_\nu \omega_\nu\right)e^{i\left[qa-2\pi\left(\sum_\nu\nu j_\nu\right)/N\right](\ell-\ell')} \prod_{\nu=1}^{N}\frac{1 }{j_\nu!(\omega_\nu)   ^{j_\nu}   } \right]
\end{align}
This gives the relation:
\begin{equation}
    C_{\ell\ell', n+1}(\omega, q) = \frac{1}{n+1}\left(\frac{q^2}{2mN}\right)\sum_\nu \frac{e^{-i 2\pi\nu(\ell-\ell')/N}}{\omega_\nu}C_{\ell\ell', n}(\omega-\omega_\nu, q) 
\end{equation}
with 
\begin{equation}
\begin{split}
    C_{\ell\ell', n=1}(\omega=\omega_\nu, q) =\frac{2\pi}{V} e^{-2W(q)} e^{iqa(\ell-\ell')}  \left(\frac{q^2}{2mN}\right)\frac{e^{-i2\pi\nu (\ell-\ell')/N} }{\omega_\nu}.  \quad 
\end{split}
\end{equation}

\section{Partial Scattering Rate for Massless Scalar Mediator}
\label{app:massless_mediator}

For completeness, here we also provide differential rates $dR/dq$ and $dR/d\omega$ for the case of a massless scalar mediator, calculated using $\sigma_p=10^{-40}$ cm$^2$. We observe similar phenomena as for the massive mediator, but in this case rates are highly peaked at low $q$ and $\omega$ due to the form factor.

\begin{figure}[h]
    \centering
    \includegraphics[width=0.99\linewidth]{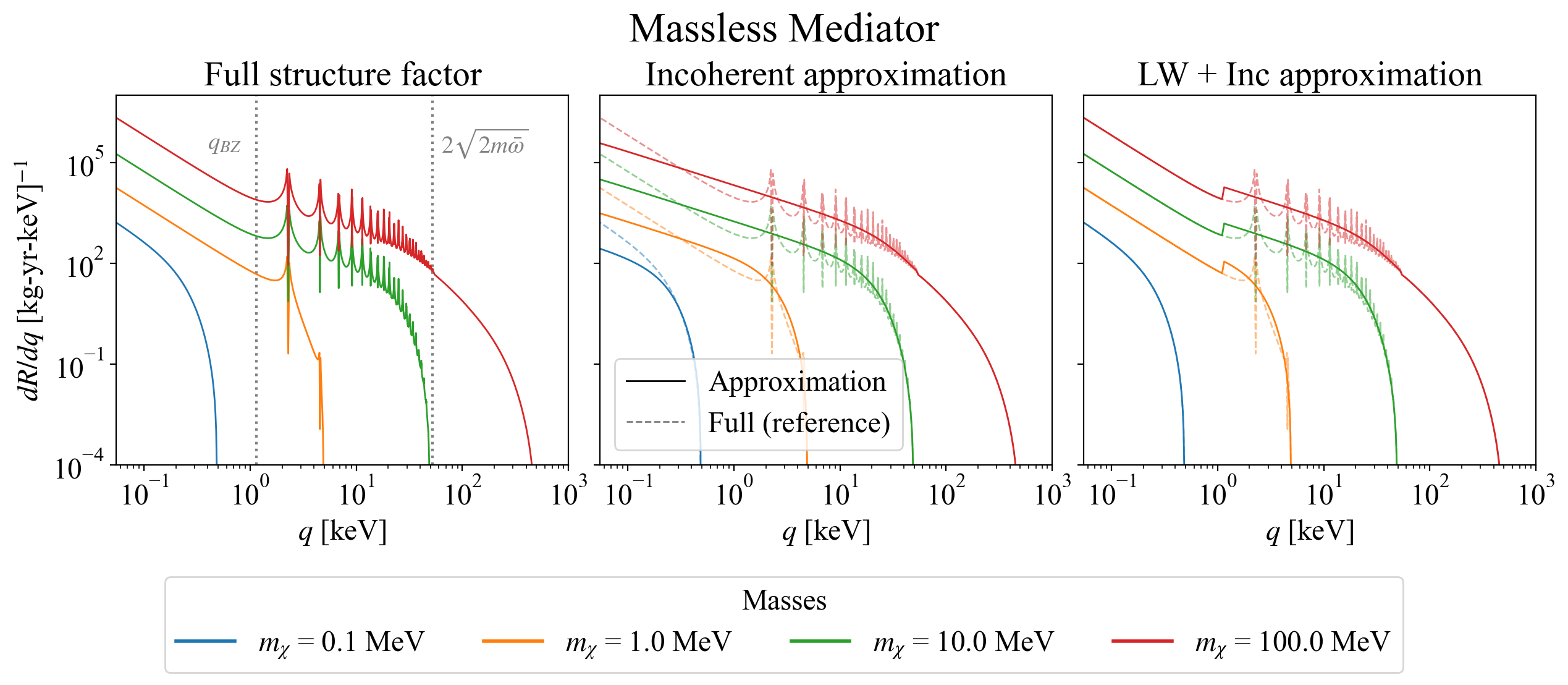}
    \caption{$dR/dq$ result for $N=100$.  }
    \label{fig:dRdq-massless}
\end{figure}

\begin{figure}[h]
    \centering
    \includegraphics[width=0.99\linewidth]{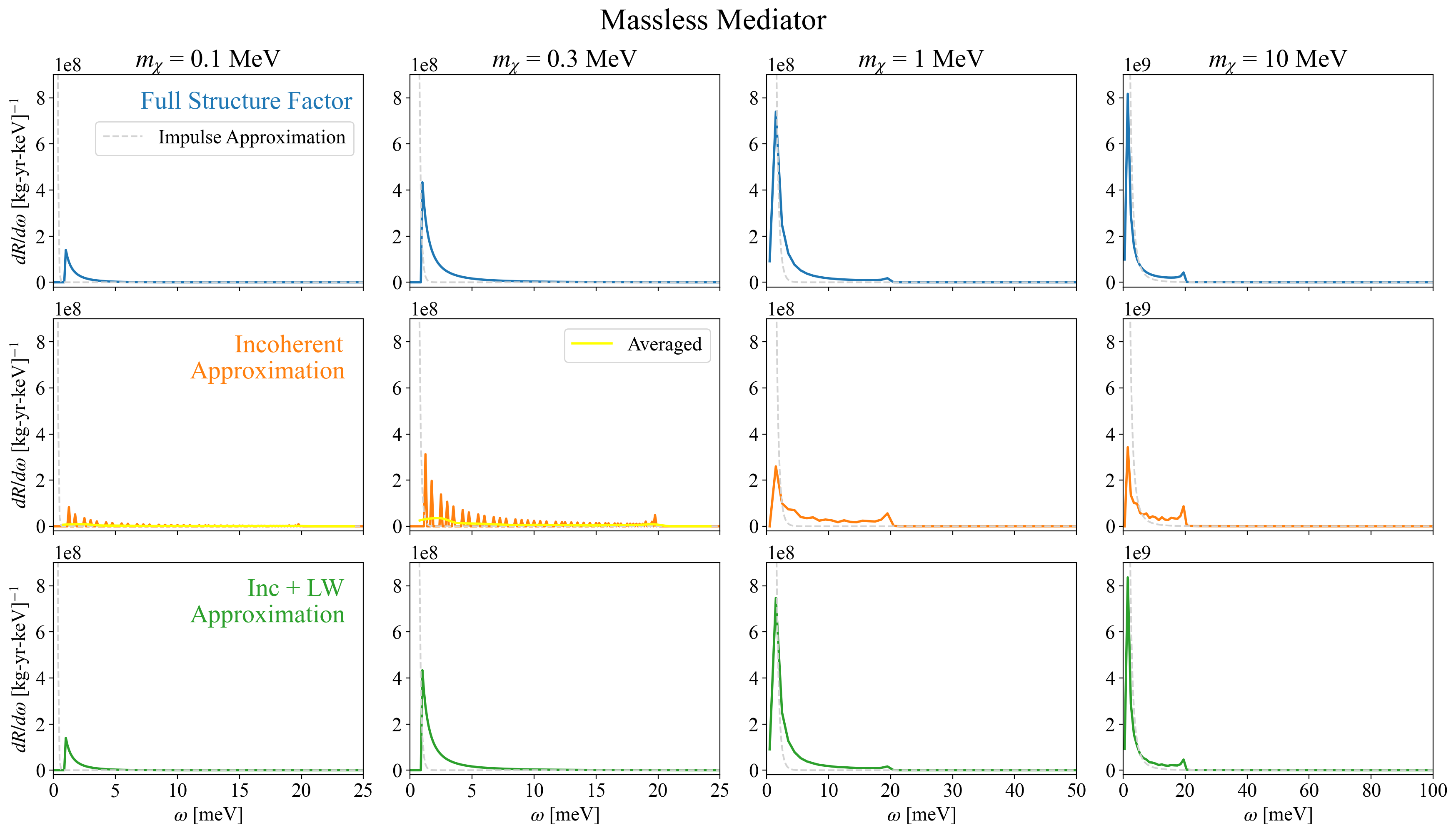}
    \caption{$dR/d\omega$ result for $N=100$.  }
    \label{fig:dRdomega-massless}
\end{figure}

\bibliographystyle{JHEP}
\bibliography{references,phonons}

\end{document}